\documentclass[12pt]{emulateapj}
\slugcomment{Accepted, to appear in ApJ, v694, 2009 April.}
\pdfoutput=1
\usepackage{wrapfig}
\usepackage{graphicx}
\usepackage{amssymb}
\usepackage{amsmath}
\usepackage{ifpdf}
\usepackage{longtable}
\usepackage{rotating}
\usepackage{natbib}
\begin{document}

\newcommand\etal{{\it et al.~}}
\newcommand\cf{{\it cf.~}}
\newcommand\eg{{\it e.g.,~}}
\newcommand\ie{{\it i.e.,~}}
\newcommand{\gtorder}{\mathrel{\raise.3ex\hbox{$>$}\mkern-14mu
             \lower0.6ex\hbox{$\sim$}}}

\title{Three-dimensional Magnetohydrodynamic Simulations of \\ Buoyant Bubbles in Galaxy Clusters}

\author{S. M. O'Neill\altaffilmark{1}, D. S. De Young \altaffilmark{2}, T. W. Jones\altaffilmark{3}}

\altaffiltext{1}{Department of Astronomy, University of Maryland, College Park, MD 20742; soneill@astro.umd.edu}
\altaffiltext{2}{National Optical Astronomy Observatory, 950 North Cherry Avenue, Tucson, AZ 85719; deyoung@noao.edu}
\altaffiltext{3}{School of Physics and Astronomy, University of Minnesota, Minneapolis, MN 55455; twj@msi.umn.edu}

\begin{abstract}
We report results of 3D MHD simulations of the dynamics of buoyant bubbles in magnetized galaxy cluster media. 
The simulations are three dimensional extensions of two dimensional calculations reported by \citet{jonesdeyoung05}. 
Initially spherical bubbles and briefly inflated spherical bubbles all with radii a few times smaller than the intracluster medium (ICM) scale height were followed as they rose through several ICM scale heights. 
Such bubbles quickly evolve into a toroidal form that, in the absence of magnetic influences, is stable against fragmentation in our simulations. 
This ring formation results from (commonly used) initial conditions that cause ICM material below the bubbles to drive upwards through the bubble, creating a vortex ring; that is, hydrostatic bubbles develop into ``smoke rings'', if they are initially {\it not} very much smaller 
or very much larger than the ICM scale height. 

Even modest ICM magnetic fields with $\beta = P_{gas}/P_{mag} \lesssim 10^3$ can influence the dynamics of the bubbles, provided the fields are not tangled on scales comparable to or smaller than the size of the bubbles. 
Quasi-uniform, horizontal fields with initial $\beta \sim 10^2$ bifurcated our bubbles before they rose more than about a scale height of the ICM, and substantially weaker fields produced clear distortions. 
These behaviors resulted from stretching and amplification of ICM fields trapped in irregularities along the top surface of the young bubbles. On
 the other hand, tangled magnetic fields with similar, modest strengths are generally less easily amplified by the bubble motions and are thus less influential in bubble evolution. 
Inclusion of a comparably strong, tangled magnetic field inside the initial bubbles had little effect on our bubble evolution, since those fields were quickly diminished through expansion of the bubble and reconnection of the initial field.
\end{abstract}

\keywords{galaxies: clusters: general -- MHD-- galaxies: active--galaxies}

\section{Introduction}
There is now ample evidence from radio and X-ray observations that active galactic nuclei (AGN) generate energetic structures that continue to interact with galaxy cluster environments even after the central engine shuts down.
Observations of detached bubbles of radio plasma and coincident X-ray cavities in cluster cores illustrate that the AGN jets inflate cocoons displacing the ambient intracluster medium (ICM) \citep[\eg][]{bohringeretal93,sleeroy98,fabianetal00,mcnamaraetal01,sleeetal01,wiseetal07}.
The $pdV$ work required to move the ICM can be estimated from X-ray observations, which suggest that upwards of $10^{59}-10^{60}~$ergs of energy are present in these cavities \citep{birzanetal04, dunnetal05,mcnamaranulsen07,wiseetal07}.
As such, they could in principle provide a reservoir of energy needed to stifle cooling flows and maintain the $\sim 2~$keV temperature floors observed in cluster cores \citep{petersonetal01,fabianetal01,kaastraetal01,tamuraetal01}.

The evolution of relic bubbles is an interesting problem in part because their structures are seen to remain coherent longer than analytic estimates would suggest.
The issue of bubble fragmentation and its associated timescale is of importance for models that use ``AGN feedback'' to reheat the ICM and thus suppress both ICM cooling flows and the occurrence of large amounts of star formation in the cluster core.  
This is because such reheating needs to be spread throughout the ICM in the central regions, and if radio bubbles do not fragment on timescales comparable to the local ICM cooling time, the distribution of their energy over large volumes requires the assumption of some other, yet unspecified mechanism.  
This fragmentation question is also related to the more general problem of AGN feedback in current cosmological models and in particular to the use of ``radio AGN feedback'' to produce the observed distribution of massive galaxy morphologies and colors \citep[\eg][]{crotonetal06}. 
In particular, because the bubbles are light, presumably being filled with very hot and possibly relativistic plasma, they buoyantly rise in the cluster potential and are subject to disruption by both the Rayleigh-Taylor (R-T) and Kelvin-Helmholtz (K-H) instabilities.
Simple linear stability analysis suggests that disruption should take 
place on timescales of $\sim 10^7~$years \citep[see][for example]{heinzchurazov05}, yet intact bubbles are seen at distances from the 
cluster core that imply rise times of  an order of magnitude longer
\citep{birzanetal04,dunnetal05}.
As \citet{deyoung03} pointed out, tension from ICM magnetic fields 
and possibly internal bubble magnetic fields could play an important role in 
relic evolution by stabilizing bubble structures against these instabilities for timescales $\sim 10^8~$years.
On the other hand, the role of a magnetic field in either of
these instabilities is dependent on both the strength and the 
orientation of the field with respect to the unstable boundary.
In addition, the field does not enhance stability for
modes orthogonal to the plane containing the field and
the boundary \citep[\eg][]{chandra,dursi07}.
Furthermore, nonlinear evolution of R-T and K-H
instabilities can be very difficult to predict, even in the
hydrodynamic case, but especially
in the presence of magnetic fields; disruptive trends can be
both enhanced and reduced compared to linear predictions,
depending on the details \citep[\eg][]{jun95,ryu00}.
Consequently, understanding
of this issue requires fully nonlinear MHD numerical simulations.
 
Several groups have previously conducted simulations in 2D and 3D to 
explore basic bubble dynamics and morphology and their relationship to 
ICM magnetic fields.
For example, \citet{churazovetal01} first established buoyant rise times
of $\sim10^8$ years for bubbles in cluster environments in 2D hydrodynamic simulations.
Hydrodynamic simulations conducted by \citet{bruggenkaiser01} in 2D 
and \citet{bruggenetal02} in 3D confirmed the anticipated
timescales of $\sim 10^7~$years for disruption from hydrodynamical instabilities in
such bubbles.
To explore the effects of magnetic fields on bubble dynamics and stability, 
\citet{robinsonetal04} and \citet{jonesdeyoung05} conducted multiple 
series of 2D magnetohydrodynamic (MHD) simulations involving
relatively homogeneous ambient magnetic fields with a variety
of field orientations with respect to the cluster gravitational
field. Once again the bubble
rise times were $\sim10^8~$years. 
These calculations also demonstrated that even ambient fields initially much too weak to provide linear stability to R-T and K-H instabilities could, nonetheless, apparently prevent bubble disruption on timescales comparable to the inferred lifetimes of the observed ``relic'' radio bubbles in clusters.
The surprising ability of such weak fields to stabilize bubble boundaries in those simulations came from the stretching of field lines in response to the bubbles' rise through the ICM and from vortical motions generated within the bubbles.
More recent work in the 3D MHD regime by \citet{ruszkowskietal07} has 
suggested, on the other hand,
that random ambient magnetic fields only provide such bubble stability if 
the coherence lengths of these fields are large with respect to the bubble size.

Magnetic fields may also play significant roles in the dynamics
and thermal conduction properties of ``cold fronts'', which
are contact discontinuities separating merging subclusters 
\citep[\eg][]{vikh01,asai07,dursi08}. That is a similar physics
problem in some respects,
although strong differences in the initial conditions, relative 
properties of the media, and the overall dynamics make
direct comparisons between magnetic field
roles in cold fronts and bubble boundaries nontrivial. We will
not address magnetic effects in cold fronts here.

In this work, we explore further the three-dimensional evolution of
bubble dynamics and morphology 
in the presence of magnetic fields.
We present results from a series of 3D MHD simulations of buoyant bubbles 
inflated in a cluster atmosphere similar to that used by \cite{jonesdeyoung05}.
We test, for several field strengths and topologies, what determines 
bubble shapes and how magnetic fields affect the evolution of bubbles.
Since the magnetic fields in the bubbles have distinct origins from
the cluster fields they encounter, we take pains in our initial 
conditions to isolate the bubble and ambient medium magnetic fields.
We also explore the dynamical evolution of buoyant bubbles that are inflated by
the injection of energetic material into an
initially small volume.
Many of the previously published calculations were based on bubbles that were
initialized as static spherical structures inside an ICM.  Since real radio bubbles
are remnants of AGN outflows and thus not likely to be particularly spherical at the
start, it is important to understand how simulation results depend on this
choice of initial conditions.

The paper is organized as follows.
Section \ref{sec:calc} describes the details of the calculations performed, including a description of the numerical algorithm and explanations of the simulated models.
Section \ref{sec:results} contains a discussion of the results and an analysis of the physics of bubble propagation and magnetic field evolution.
Conclusions and a summary are provided in $\S$ \ref{sec:conc}.

\begin{deluxetable*}{cccccc}
\label{table}
\tabletypesize{\scriptsize}
\tablecaption{Summary of Simulations of Buoyant Bubbles}
\tablewidth{0.7\textwidth}
\tablehead{
\colhead{ID \tablenotemark{1}} &
\colhead{ICM Field Geometry} &
\colhead{$\beta_a$ \tablenotemark{2}} &
\colhead{$B_a$ ($\mu$G) \tablenotemark{2} \tablenotemark{3}} &
\colhead{$t_i$ (Myr)} &
\colhead{$t_e$ (Myr)} \\}
\startdata
US   & Uniform & 120    & 5 & 0 & 150\\
US-I & Uniform & 120    & 5 & 10 & 136\\
US-F & Uniform & 120    & 5 & 0 & 150\\ 
UM   & Uniform & 3000   & 1 & 0 & 150\\
UM-I & Uniform & 3000   & 1 & 10 & 114\\
UW   & Uniform & $\infty$ & 0 & 0 & 150\\
UW-I & Uniform & $\infty$ & 0 & 10 & 132\\
T$_L$S   & Tangled-L & 120    & 5 & 0 & 108\\
T$_L$S-I & Tangled-L & 120    & 5 & 10 & 114\\
T$_L$S-F & Tangled-L & 120    & 5 & 0 & 108\\
T$_B$S   & Tangled-B & 120    & 5 & 0 & 114\\
T$_B$S-I & Tangled-B & 120    & 5 & 10 & 130\\
\enddata
\tablenotetext{1}{The first letter indicates the ambient magnetic field geometry: uniform (U) or tangled (T) with coherence lengths that are long (sub L) or bubble-sized (sub B).  
The second letter refers to the strength of the field: Strong (S), Moderate (M), or Weak (W).
Letters to the right of the dash indicate either that the bubble is inflated (I) or that the bubble itself contains a magnetic field (F), while models without these indicators imply the opposite case (\ie non-inflated or without a bubble field).
}
\tablenotetext{2}{This is only true in a statistical sense for the tangled fields.}
\tablenotetext{3}{Measured at $x=30~$kpc.}
\end{deluxetable*}

\section{Calculation Details}\label{sec:calc}
The computational approach we describe here is similar to that of \citet{jonesdeyoung05} but extended to three dimensions.
We inflated bubbles from a spherical volume of
radius $r_b$ near the base of a plane stratified ICM atmosphere initially in
hydrostatic equilibrium. 

In most of the simulations the ICM was magnetized with magnetic field pressures that were generally much smaller than the ICM thermal pressure.
In one set of simulations, the ambient magnetic field was quasi-uniform, meaning that the field asymptotically achieves uniform orientation far from the bubble inflation region.
In another set of models, the ICM field was tangled on scales either comparable to or larger than the bubble inflation region.
Cases are included with magnetic field strengths interior to the inflated bubbles comparable to the external fields and with bubble fields absent.
For comparison, we also include two strictly hydrodynamic cases without any ICM magnetic fields.
General simulation descriptions follow, with specific details of the ICM given in $\S$ \ref{sec:icm} and those of the bubbles in $\S$ \ref{sec:bubbles}. 

Our simulations were conducted on a 3D Cartesian grid of total 
physical dimensions 
$x_{size} = 40 {\rm~kpc}$, $y_{size} = z_{size} = 34 {\rm~kpc}$, 
covering the ranges $x = [5 {\rm~kpc}: 45 {\rm~kpc}]$, $y = z \approx [-17 {\rm~kpc}: 17 {\rm~kpc}]$.
The gravitational acceleration is in the $-x$ direction in this coordinate system.
The bubble origin was centered at $x_b = 13 {\rm~kpc}$, $y_b = z_b = 0$. 
The grid was partitioned into uniform computational zones of size $\Delta x = \Delta y = \Delta z = 133 {\rm~pc}$.
At the $y$ and $z$ boundaries, continuous boundary conditions were maintained.
To maintain hydrostatic equilibrium in the $x$ direction, extrapolation boundaries were employed at $x=5$ and $x=45~$kpc.
In particular, the density in a given boundary zone was set through a linear extrapolation of the densities in the adjacent pair of physical grid zones.
Using this updated boundary density, the boundary pressure was derived assuming hydrostatic equilibrium within the boundary zones.
Additionally, the velocities were chosen to be continuous across both $x$ boundaries, in order to minimize the impact of waves incident upon those zones. 
The calculations were terminated when any bubble material approached any boundary of the computational domain (generally $x_{top}$).

Our simulations employed a second-order, nonrelativistic, Eulerian, total variation 
diminishing (TVD), ideal 3D MHD code, described in \citet{ryuj95} and \citet{ryuetal98}.
The code explicitly enforces the divergence-free condition for magnetic fields through a constrained transport scheme detailed in \citet{ryuetal98}.
The numerical method conserves mass, momentum, and energy to machine accuracy.
To set up hydrostatic equilibrium in the ambient medium in the
simulations presented here, acceleration due to the assumed gravity was 
added as a source 
term in the $-x$ direction through operator splitting. The corrected
$x$ momentum at each timestep was used to recalculate the total kinetic
plus thermal and magnetic energy.
In this simplified treatment of gravity, momentum and energy are no longer 
exactly conserved, but
we have verified that associated errors are much too small to influence our results.
A gamma-law gas equation of state was assumed with $\gamma = 5/3$.
A passive mass fraction or ``color'' tracer, $C_f$, was introduced for material originating inside the bubble inflation region.
$C_f$ was set to unity inside the bubble, while $C_f = 0$ in the ambient medium.

\subsection{Gravity and The Ambient Medium}\label{sec:icm}
We applied the same gravity model as \citet{jonesdeyoung05}, which
was derived from the combined mass distributions of the central active 
galaxy and the cluster.
The cluster had an NFW mass profile, while the active galaxy had
a King model.
The galaxy and cluster mass normalizations were chosen such that 
the cluster mass inside 10 kpc was $3.5 \times 10^{10} M_{\sun}$
and the galaxy mass inside 20 kpc was 
$3.5 \times 10^{12} M_{\sun}$.
Inside $r = 60~$kpc (i.e., everywhere on our computational
domain) matter associated with the active galaxy dominated the
gravitational force.

\begin{figure*}[t]
\includegraphics[width=\textwidth]{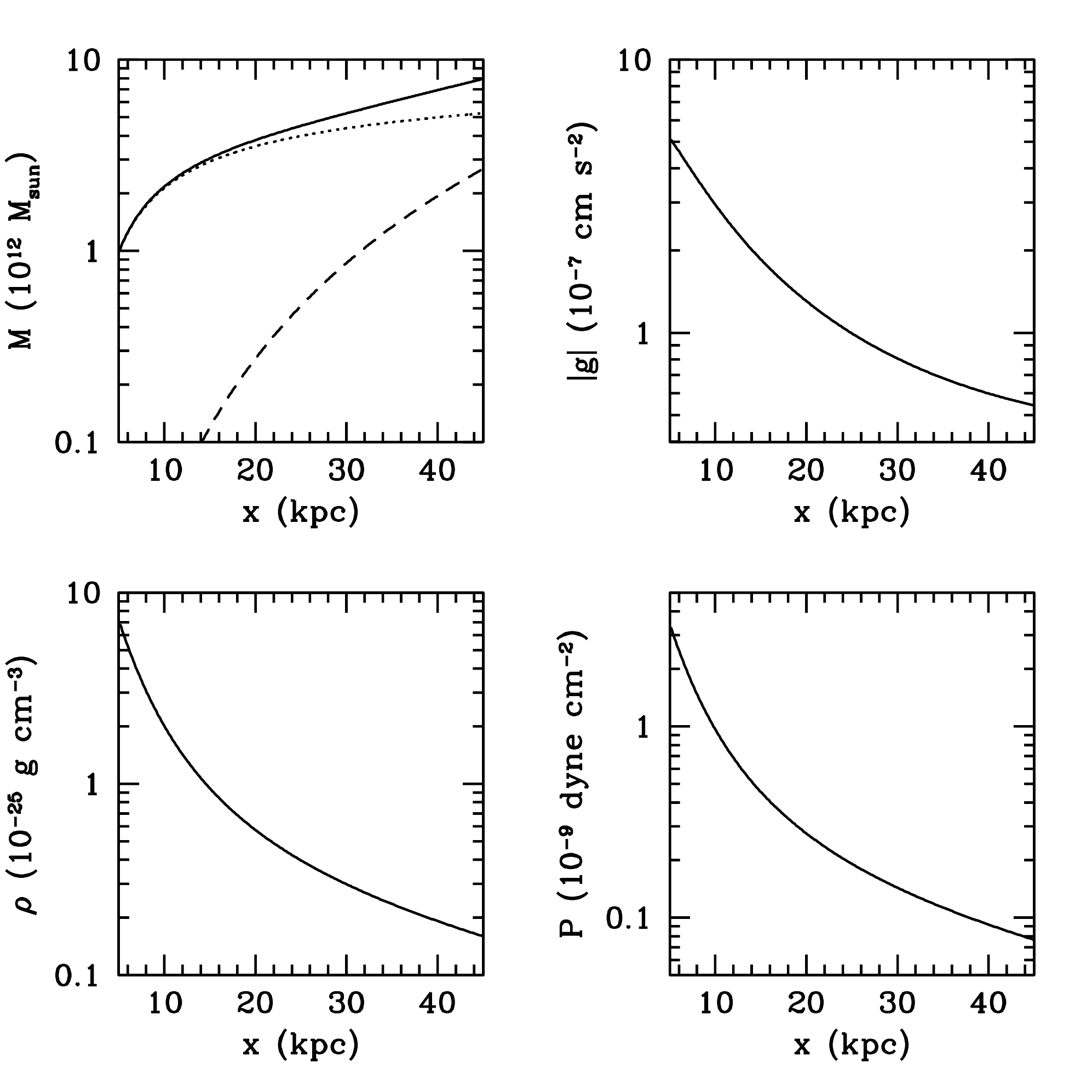}
\caption[Initial buoyant bubble environments]{Initial buoyant bubble environments. {\it Top left:} Gravitational mass as a function of radius from the core: total (solid line), galaxy (dotted line), and cluster (dashed line).  {\it Top right:} Gravitational acceleration.  {\it Bottom left:} Gas density.  {\it Bottom right:} Gas pressure.}
%fig 1
\label{fig:profiles}
\end{figure*}

To match the planar ICM geometry of \citet{jonesdeyoung05}, we projected
the resulting gravitational acceleration onto the $x$ direction;
that is, the gravity was one dimensional, leading to a
plane stratified hydrostatic equilibrium for the initial ICM.
That ICM was assumed to be isothermal with 
$T_0 \approx 3.5 \times 10^7~$K, or, equivalently, $kT_0 = 3~$keV.
We assumed a hydrogen mass fraction $X=0.75$ 
(for a mean molecular weight $\mu = 0.6$), giving an adiabatic ICM 
sound speed $c_{s,0} = \sqrt{\gamma P_0/\rho_0} = 894~{\rm km~s}^{-1} = 0.9~{\rm kpc~Myr}^{-1}$.
The electron density was set to  $n_e=0.1~{\rm cm}^{-3}$ at $x=5~$kpc,
leading to the atmospheric density, $\rho_0$, and pressure, $P_0$, profiles shown in 
Figure \ref{fig:profiles}. 
They are identical to those in the \citet{jonesdeyoung05} simulations.
The scale height of the ICM, here defined to be $h = c_{s,0}^2/|g|$, 
turns out to behave as $h \approx x$, which allows us to approximate
the density and pressure distributions as a simple power law
for analysis purposes; in particular
\begin{equation}
P_0(x) \approx \frac{c^2_{s,0}}{\gamma}\rho_0(x) = P_{x_0}\left(\frac{x}{x_0}\right)^{-\gamma},
\end{equation}
where $\gamma = 5/3$ is the gas adiabatic index, $x_0$ is some fiducial vertical position, and $P_{x_0}$ is the pressure at $x_0$.

\begin{figure*}[t]
\begin{center}
\includegraphics[width=0.8\textwidth]{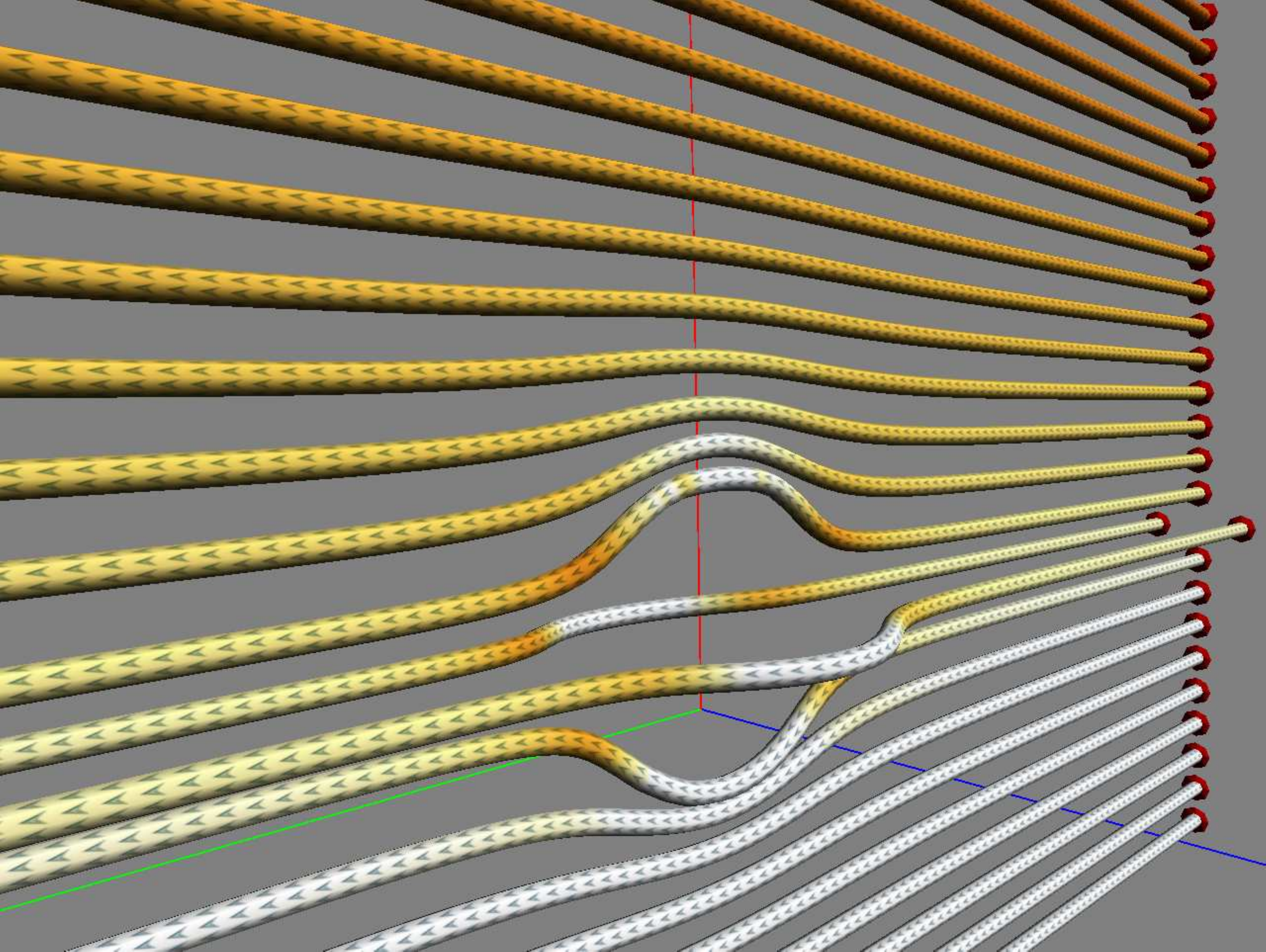}
\caption[Magnetic field configuration used in the uniform field models]{Magnetic field configuration used in the uniform (``U'') field models.  Line brightness is indicative of magnetic field strength.  The field lines bend to accommodate the bubble inflation region near the middle of the image.  Once initialized, a buoyant bubble would propagate from the initialization region toward the top of the image.}
%fig 2
\label{fig:unifields}
\end{center}
\end{figure*}

We explored consequences of two simple ICM magnetic field models,
one quasi-uniform and one tangled, with
both designed so that the ICM field did not initially penetrate
into the bubble. 
The quasi-uniform ICM field, which we identify using the letter ``U''
in model labels, was a
3D extension of the 2D field model used by
\citet{jonesdeyoung05}. In this case field lines initially lay in the $x-y$ plane, being
derived from a vector potential $A_z(r,\phi) = B_0 r[1-(r_b(z)/r)^2]\cos{\phi}$, defined in cylindrical coordinates $(r,\phi,z)$ with respect to the center of the bubble
inflation region with $\phi = 0$ along the $\hat x$ direction (\ie the direction of bubble propagation).
In this expression $r_b(z) = r_b\sqrt{1 - ((z-z_b)/r_b)^2}$ for
$|z-r_b|\le r_b$.
At large distances from the bubble, the field lines in this model become horizontal, but they are
tangent to the bubble surface when $r = r_b(z)$, so that they skirt 
over its top and bottom in the vertical plane.
Figure \ref{fig:unifields} shows the geometry of the quasi-uniform ICM field. 
The field strength parameter $B_0$ was scaled vertically
as $B_0(x) \propto \sqrt{P_0(x)}$, so that the ratio of
gas to magnetic pressures, $\beta_0 \equiv 8\pi P_0/B_0^2$ was constant 
with height away from the inflation region.
Thus, the ambient magnetic field strength generally decreased with
height in the ICM.

To provide a direct comparison to the 2D models of 
\citet{jonesdeyoung05} we
carried out in the present study several quasi-uniform,  ``U'', field model simulations with
the same values of $\beta_0$ that they chose; namely,
$\beta_0 = 120$ (tagged ``S'' for strong field), $\beta_0 = 3000$ (tagged
``M'' for medium field) and $\beta_0 = \infty$ (tagged ``W'' for
weak field). These ISM field models are identified by the codes
``US'', ``UM'' and ``UW'' in Table 1.

\begin{figure*}[t]
\begin{center}
\includegraphics[width=0.45\textwidth]{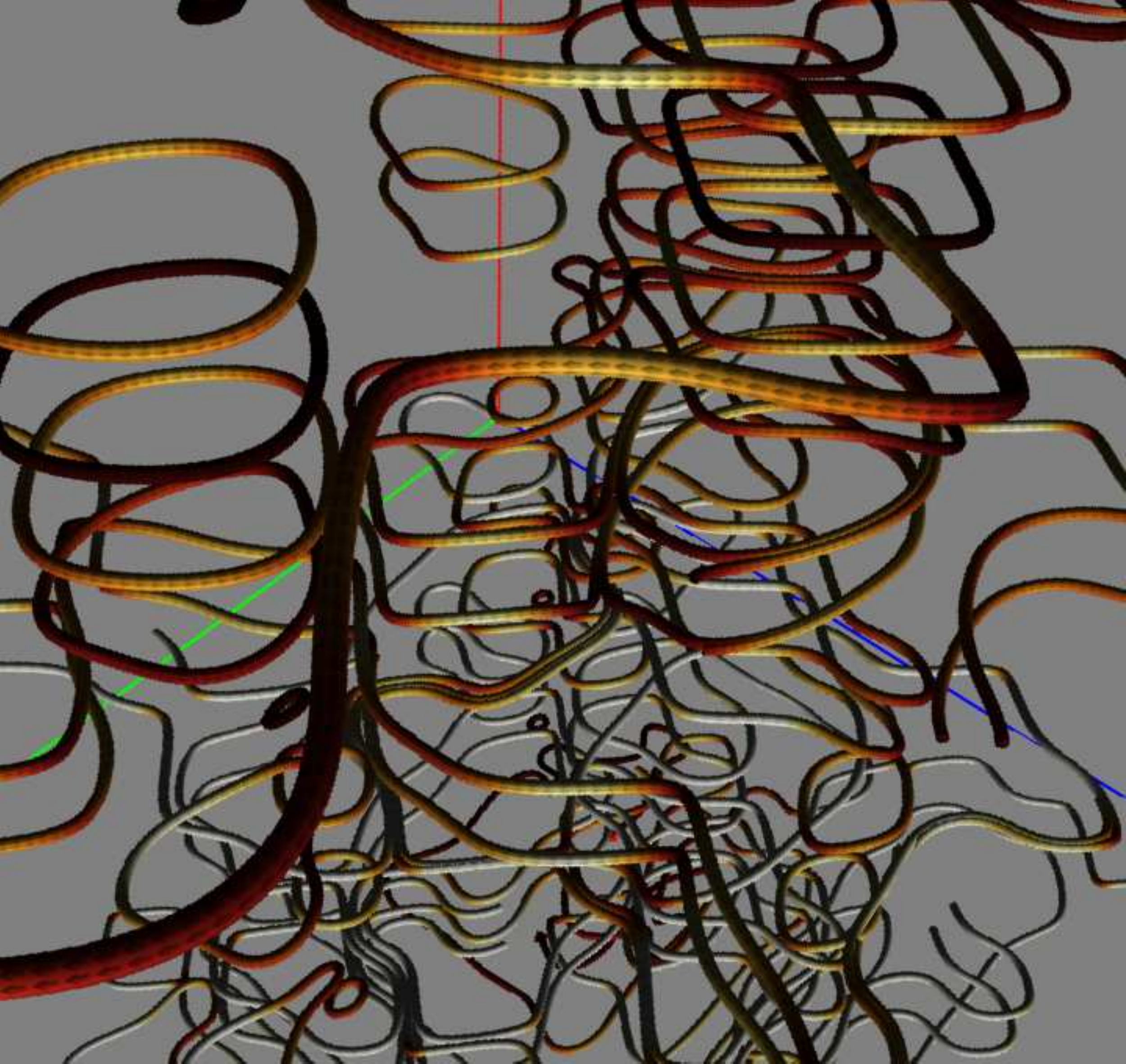}
\includegraphics[width=0.45\textwidth]{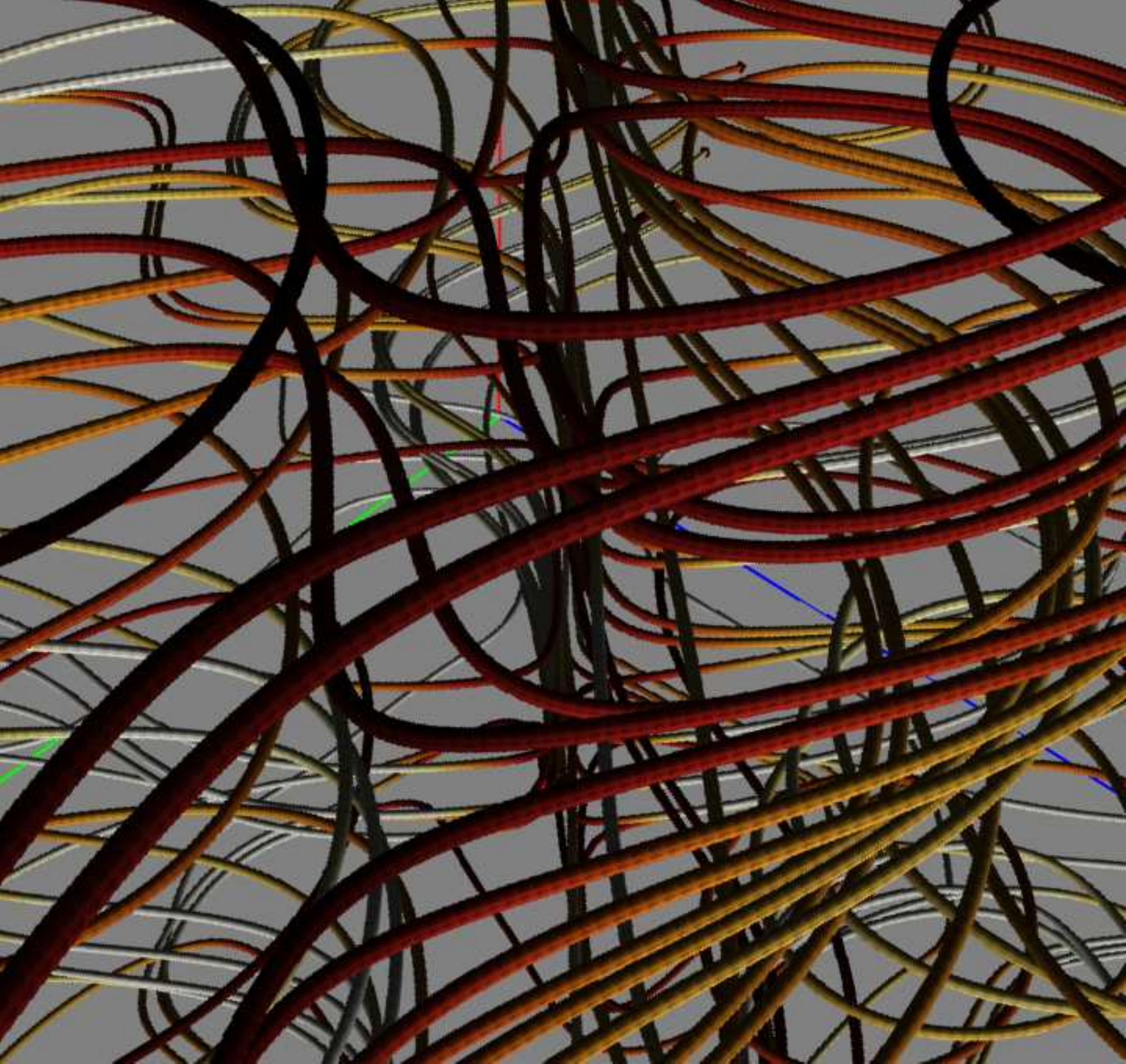}
\caption[Magnetic field configurations used in the tangled field models]{Magnetic field configurations used in the T$_B$ (left) and T$_L$ (right) models.  Line brightness is indicative of magnetic field strength.  Once initialized near the very bottom of the image, a buoyant bubble would propagate from the initialization region toward the top of the image.}
%fig 3
\label{fig:tangledfields}
\end{center}
\end{figure*}

In reporting their 3D simulation results, \citet{ruszkowskietal07}
emphasized that the stabilizing influence of their turbulent ICM 
magnetic field was effective only if the field coherence length
exceeded the size of a bubble. In order to examine the dependence
of that result on the details of the field model, we also
carried out several simulations with a simple tangled ICM
field model, which we identify with the label ``T''. 
While admittedly somewhat unrealistic, the
model we employed has the advantage of having a simple, well-defined 
coherence length that enables us to
test directly the importance of this one parameter in
determining the dynamical role of the field.
In this model the field lines initially lie in the {\it horizontal}, $y-z$
plane with
\begin{equation}
\begin{aligned}
B_x &= 0 \\
B_y &= B_0(x) f(x,z)[\sin(k_{x1}x)\sin(k_{z1}z) + \cos(k_{x2}x)\cos(k_{z2}z)] \\
B_z &= B_0(x) f(x,y)[\sin(k_{x1}x)\sin(k_{y1}y) + \cos(k_{x2}x)\cos(k_{y2}y)]. \\
\end{aligned}
\end{equation}
In these expressions, the $x$, $y$, and $z$ values are measured with respect to the grid origin rather than the bubble origin.
Far from the bubble $f \rightarrow 1$, while $f$ decreases exponentially
towards zero at the boundary of the bubble inflation region, so that
this ICM field once again does not initially penetrate into the
bubble.
The field structure-scale parameters, $k_{i,j}$ ( i = x,y,z; j = 1,2) were chosen either to make structures large compared to the bubble inflation region ($k_{i,j} < \pi/r_b$) (leading to a model tag ``${\rm T_L}$'' for ``tangled on large scales''), or comparable ($k_{i,j} \sim \pi/r_b$) (leading to a model tag ``${\rm T_B}$'' for ``tangled on a bubble scale'').
The actual numbers were $k_{x1} = \pi/(1.2r_b)$ and $k_{x2} = \pi/(0.7r_b)$ for the ``$T_B$'' 
models and $k_{x1} = \pi/(6.0r_b)$ and $k_{x2} = \pi /(3.5r_b)$ for the ``$T_L$'' models.
The remaining parameters were set so that $k_{i,j} = (1+\delta) k_{x,j}$ ( i=y,z; j = 1,2), where $\delta = \pm 0.04$, providing similar scales of field variation.
In this case we included only a nominally strong field, $\beta_0 = 8\pi P_0(x)/B_0^2(x) \approx 120$.
Note in this field model that the local magnetic pressure
varies substantially within horizontal planes of constant $P_0(x)$, so 
local variations in $\beta$ are large.
To quantify this, we empirically measured the average value of $\beta$ in the ICM and found that $\bar{\beta} \approx 130$ with a standard deviation of comparable magnitude.
The characteristic field strength once again decreased with height.
These (strong field) ICM models are identified in Table 1 by the labels
``${\rm T_LS}$'' and ``${\rm T_BS}$''.
Figure \ref{fig:tangledfields} graphically illustrates the two tangled field structures. 

\subsection{The Bubbles}\label{sec:bubbles}

While it has been common to conduct bubble stability studies 
beginning with static, pre-formed bubbles \citep[\eg][]{robinsonetal04,ruszkowskietal07},
instead \citet{jonesdeyoung05} gently ``pressure inflated'' their bubble structures 
over a period of at least 10 Myr during their simulations. 
Since real relic bubbles are not created instantaneously, they argued
that relevant stability issues, including magnetic field
geometries, may be sensitive to that fact.
As a simple compromise between static bubbles and full MHD jet
simulations to create bubbles, they fixed the
density and pressure inside a localized ``inflation region'',
allowing their bubbles to expand subsonically into the ICM
over the inflation time period, $t_i$. At the end of that time
the inflation region was allowed to relax
naturally within the surrounding flow. We have
extended that scheme to 3D for five of the computations
presented here, and, for direct comparison, also carried out seven
simulations with full-sized spherical bubbles pre-formed 
in the same ICM conditions as for the inflated bubbles.

The bubble inflation region in the simulations reported here
consisted of a sphere of radius 
$r_b = 2 {\rm~kpc}$, centered at $x_b = 13 {\rm~kpc}$, $y = z = 0$.
Conditions in this region were kept fixed for times $t\le t_i$
using density, $\rho_b = \eta \rho_0(x_b)$, with $\eta = 0.01$,
and twice ambient pressure ($P_b = 2 P_0(x_b)$).
These conditions were
relaxed once $t $ reached $ t_i$, so that the inflation
region was allowed to evolve naturally for $t > t_i$.
The passive $C_f$ tracer was set to unity inside this region when 
inflation was underway, so that during subsequent evolution
gas originating with bubble inflation has $C_f = 1$.
We either set $t_i = 0$ to simulate pre-formed
bubbles or $t_i = 10$ Myr for inflated bubbles, which we 
identify using an added model tag ``-I''.
We note that all of the bubble simulations reported here involved
inflation times $t_i$ shorter than the time necessary for
the inflating bubble to expand to the size of the local ICM scale
height, which is about 25 Myr in this case. As \cite{jonesdeyoung05} pointed out,
when bubble inflation is maintained longer than this, the rising
bubble creates a crude de Laval nozzle in the ICM that expels bubble material
at mildly supersonic speeds upward into the ICM. 
This results in the formation of a 
plume structure evident in both 2D \citep{jonesdeyoung05} and 3D \citep{bruggenetal02} simulations.
If this plume does not develop, then initially spherical
bubbles comparable in size to the local ICM scale height
tend to evolve rapidly into a torus around a vertical axis
before disruptive instabilities
can develop. We will discuss this behavior in \S 3.1.

In most of the simulations reported here the bubble inflation region
itself was unmagnetized.
However, in two models, denoted  by ``-F'', we incorporated a tangled 
internal field derived from the vector potential
\begin{equation}
A_r(r,\theta,\phi) = \frac{1}{m} B_b(r) r \sin^2{ \theta} \sin{(m \phi)},
\end{equation}
defined in spherical coordinates centered at the origin of the
bubble inflation region with the polar axis along $\hat z$ and
$\phi = 0$ along $\hat x$, the direction of bubble propagation.
This leads to the circumferential field
\begin{equation}
\vec{B}=B_\theta \hat{\theta} + B_\phi \hat{\phi}
\label{eq:field1}
\end{equation}
\noindent with components
\begin{equation}
B_\theta = B_b(r) \sin {\theta} \cos (m\phi)
\label{eq:field2}
\end{equation}

\noindent and

\begin{equation}
B_\phi = - \frac{1}{m} B_b (r) \sin(2\theta)\sin(m \phi),
\label{eq:field3}
\end{equation}
where $m$ is an integer.
In our simulations $m = 2$, to allow for multiple full field reversals over the azimuthal direction.
This field has zero magnetic helicity, since everywhere $\vec{A}\cdot \vec{B} = 0$.
We set $B_b(r = 0) = 0$ and in the two ``-F'' models 
increased it linearly to $B_b(r = 0.8 r_b) = B_0(x_b)$, whereupon it 
decreased quadratically to zero at $r = r_b$. 
This maximum value $B_0(x_b)$ is approximately four times the average ICM field strength and more than twice the peak ICM field strength, each measured at $x=x_b$.

\section{Results}\label{sec:results}

\begin{figure*}
\includegraphics[width=0.5\textwidth]{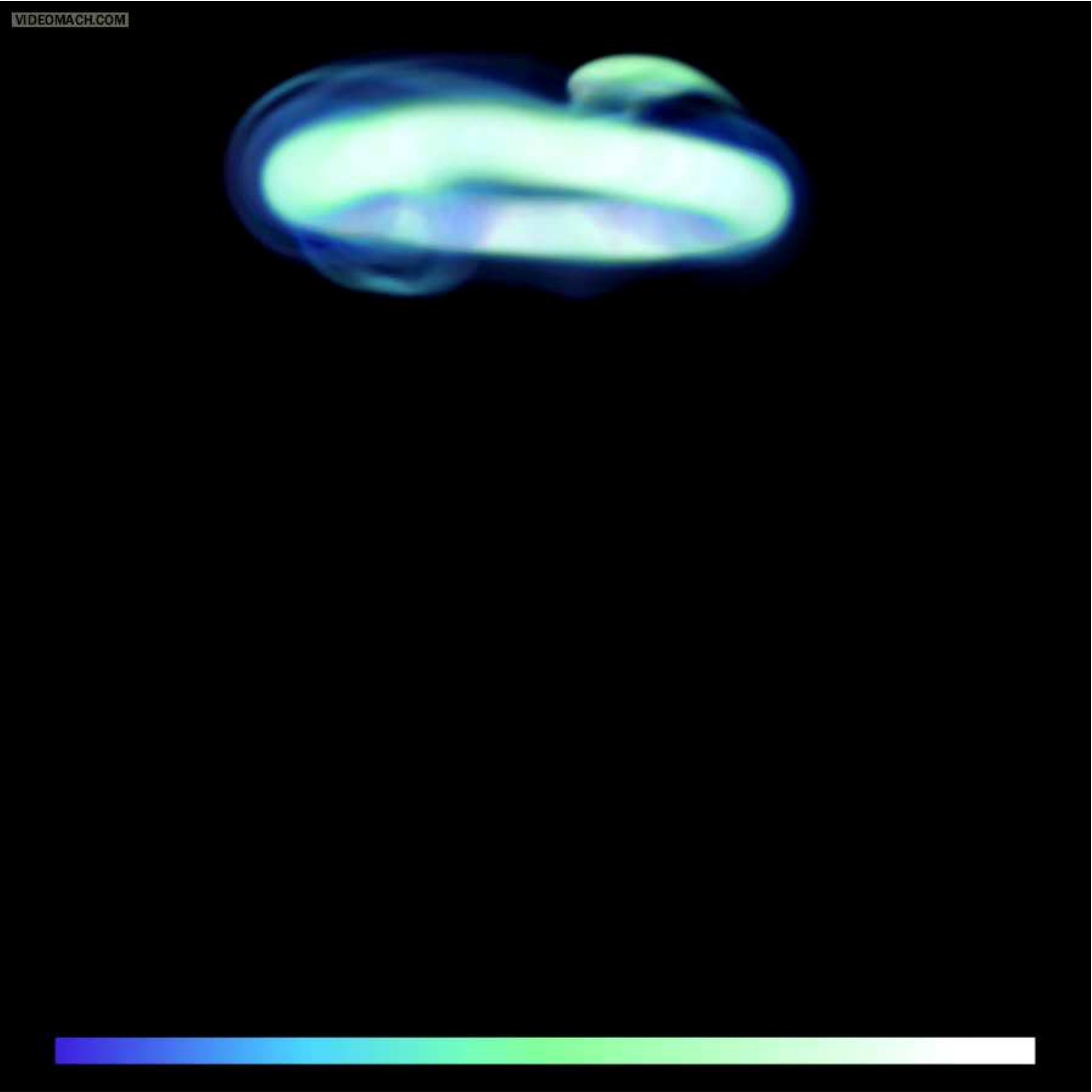}
\includegraphics[width=0.5\textwidth]{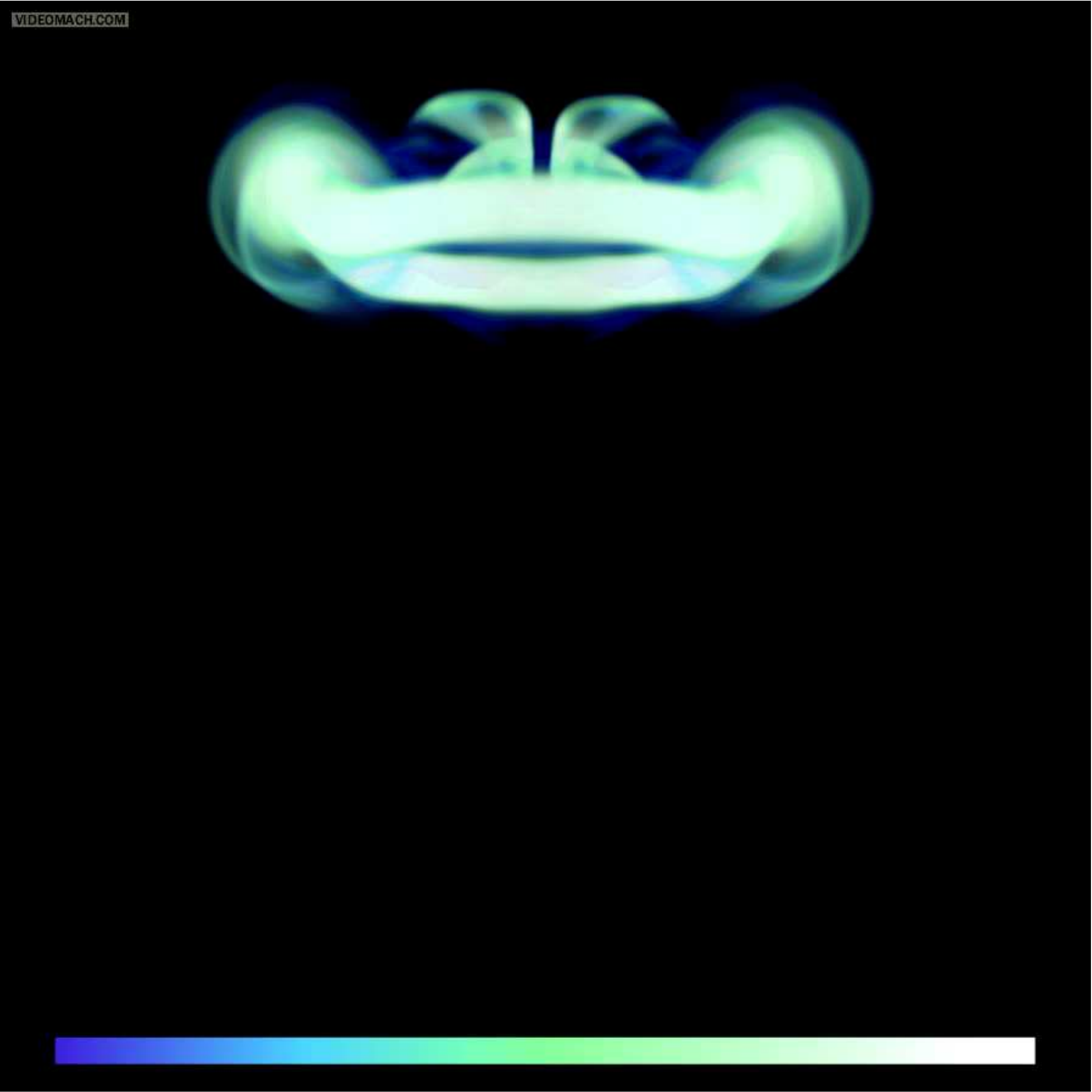}
\includegraphics[width=0.5\textwidth]{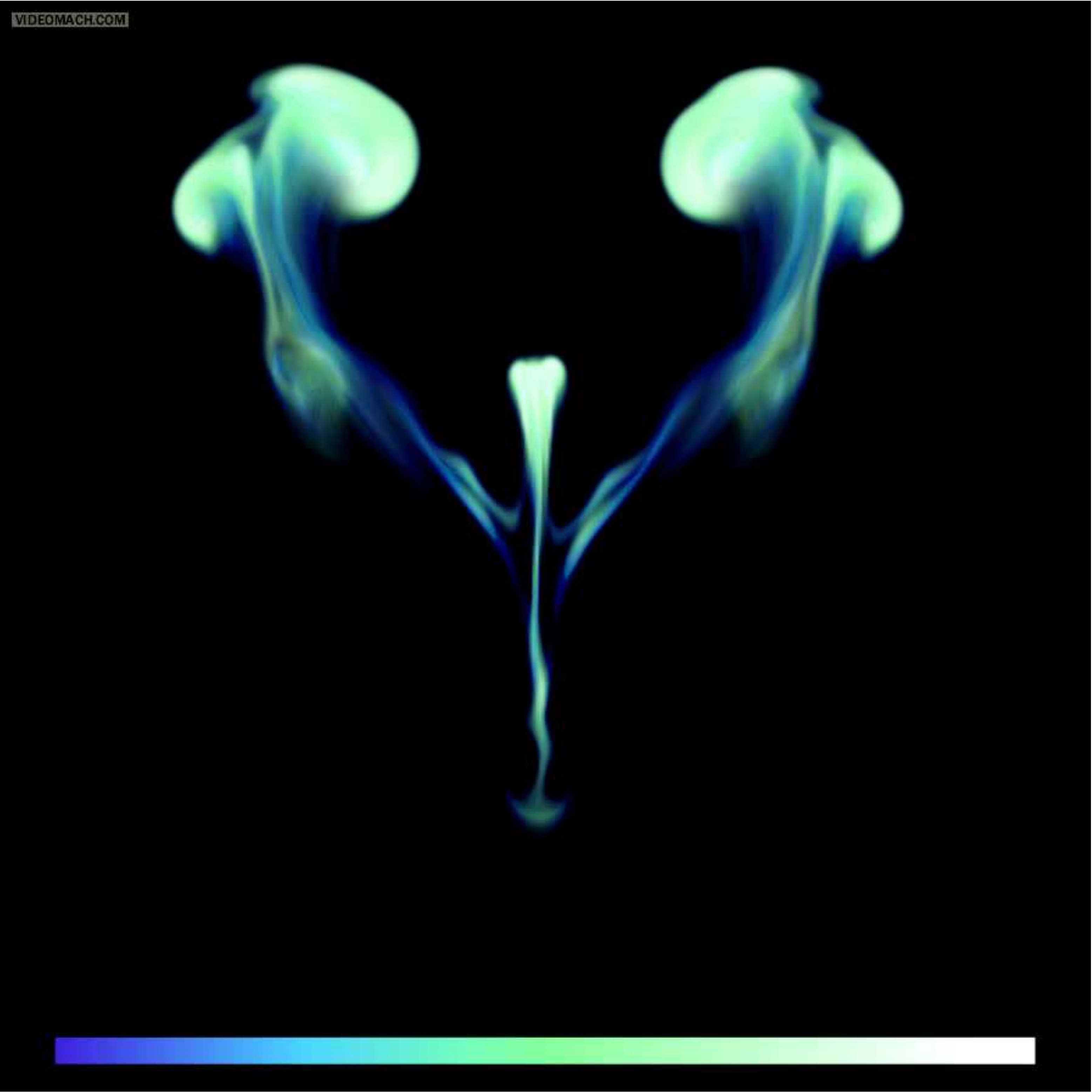}
\includegraphics[width=0.5\textwidth]{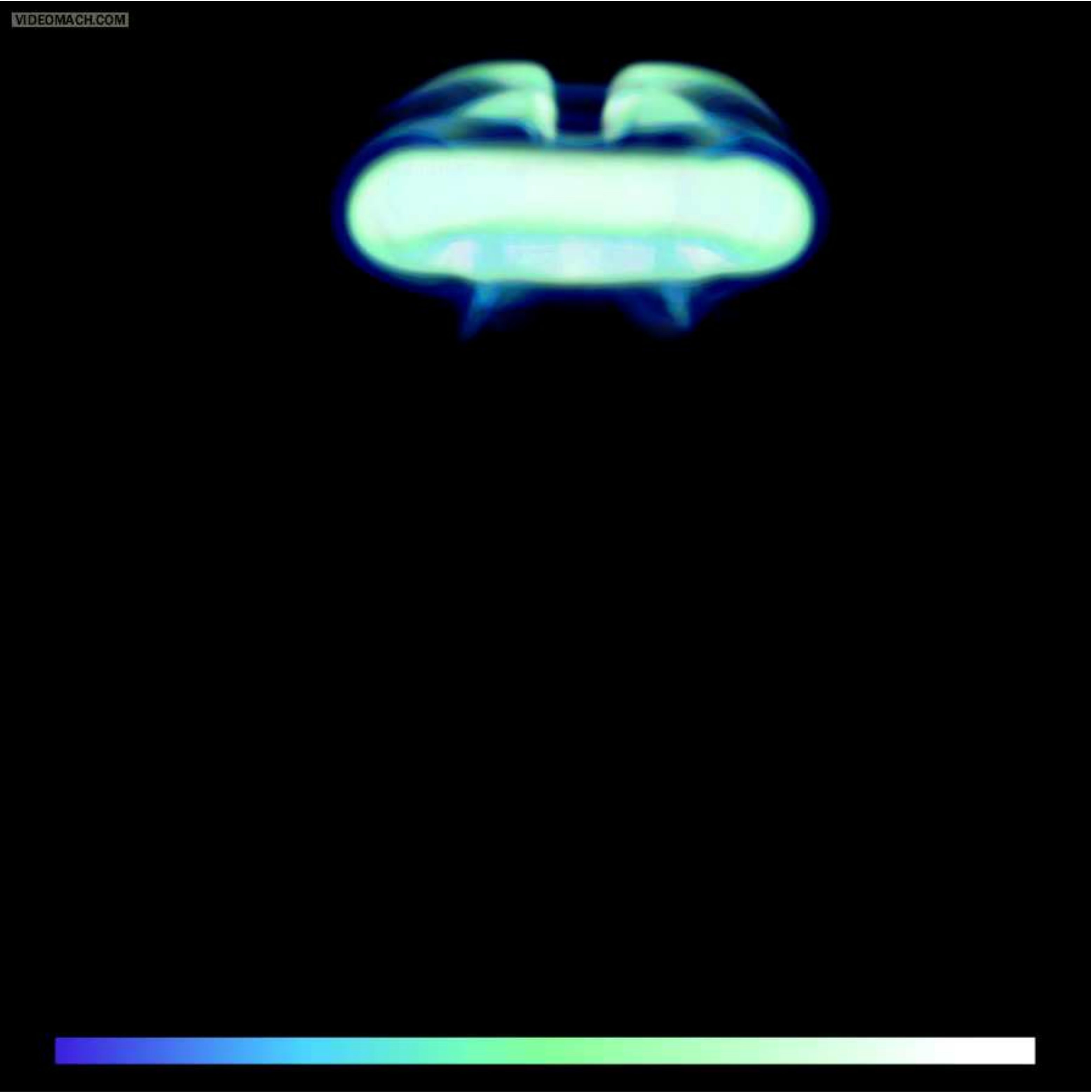}
\caption[Volume renderings looking along the $y$ direction of the 
passive ``color'' variable for inflated bubbles]{Volume renderings of 
the passive ``color'' variable for inflated bubbles near the end of 
their evolution, where light colors represent bubble material and dark 
represent the ICM.  The ambient magnetic field is along the line of 
sight.  Animations of these quantities as seen from several different angles are available at http://www.astro.umn.edu/groups/compastro/ under ``Buoyant Bubbles in Galaxy Clusters.'' 
%fig 4
{\it Top left:} UW-I model. {\it Top right:} UM-I model. 
{\it Bottom left:} US-I model.  {\it Bottom right:} T$_B$S-I model.}
\label{fig:xzinfrings}
\end{figure*}

\begin{figure*}
\includegraphics[width=0.5\textwidth]{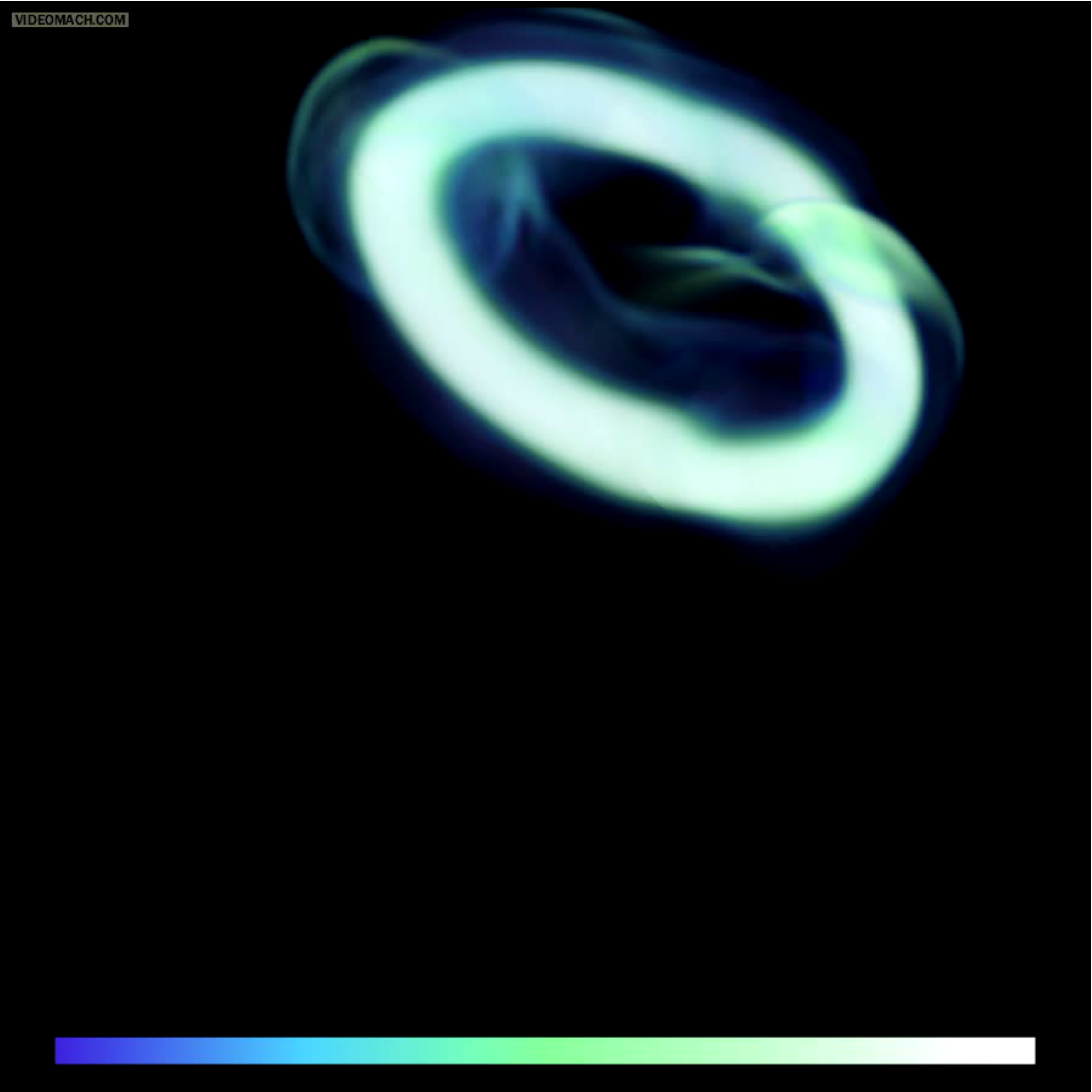}
\includegraphics[width=0.5\textwidth]{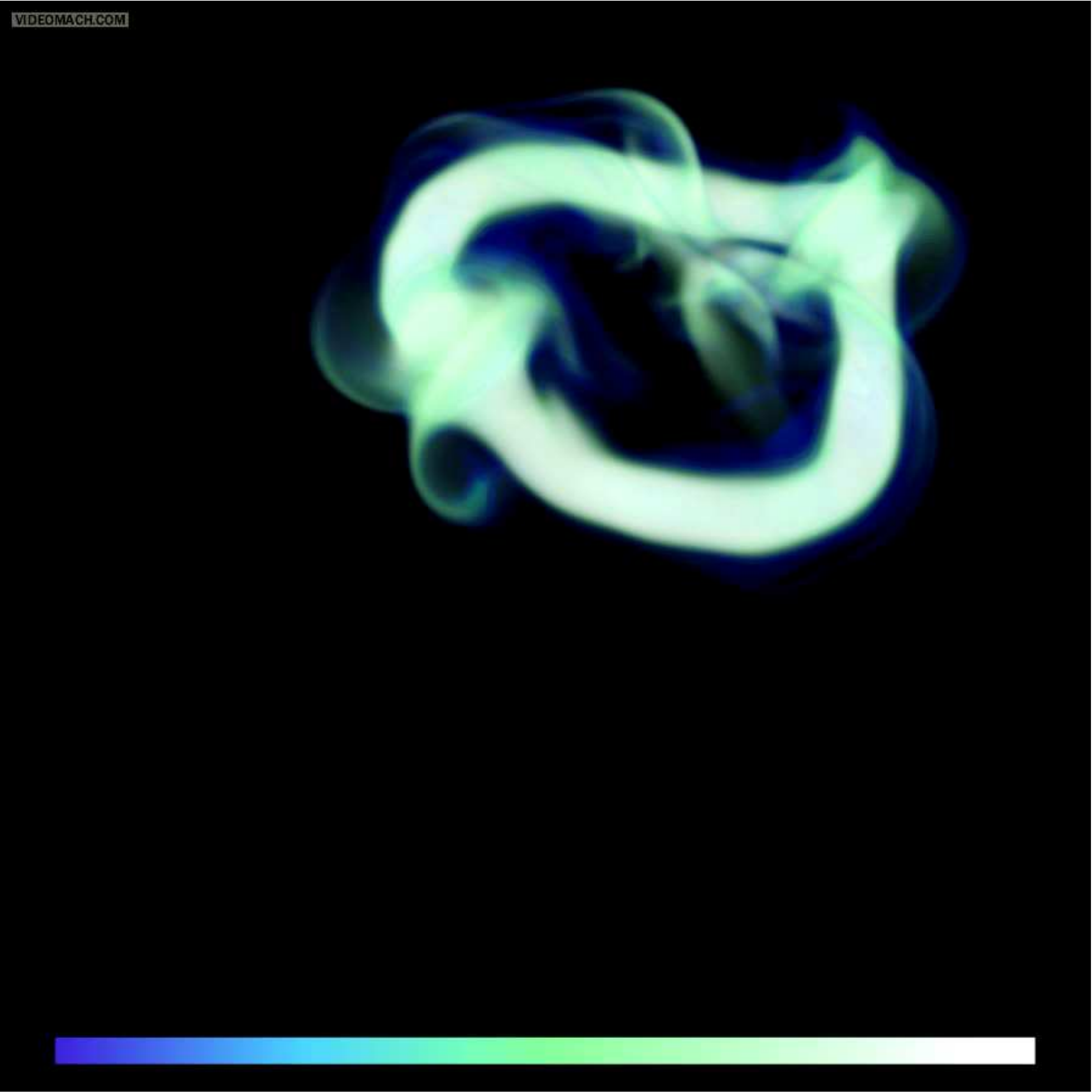}
\includegraphics[width=0.5\textwidth]{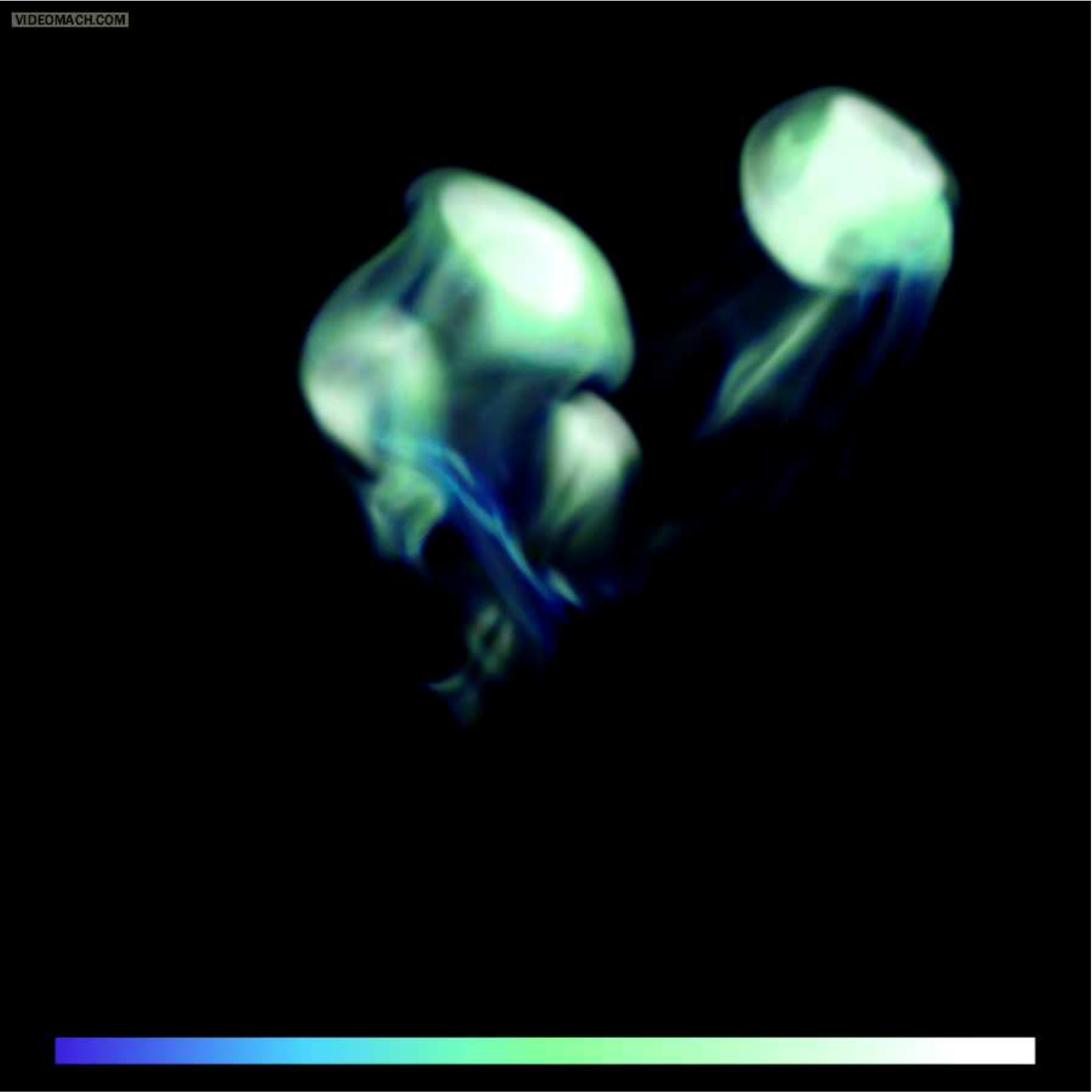}
\includegraphics[width=0.5\textwidth]{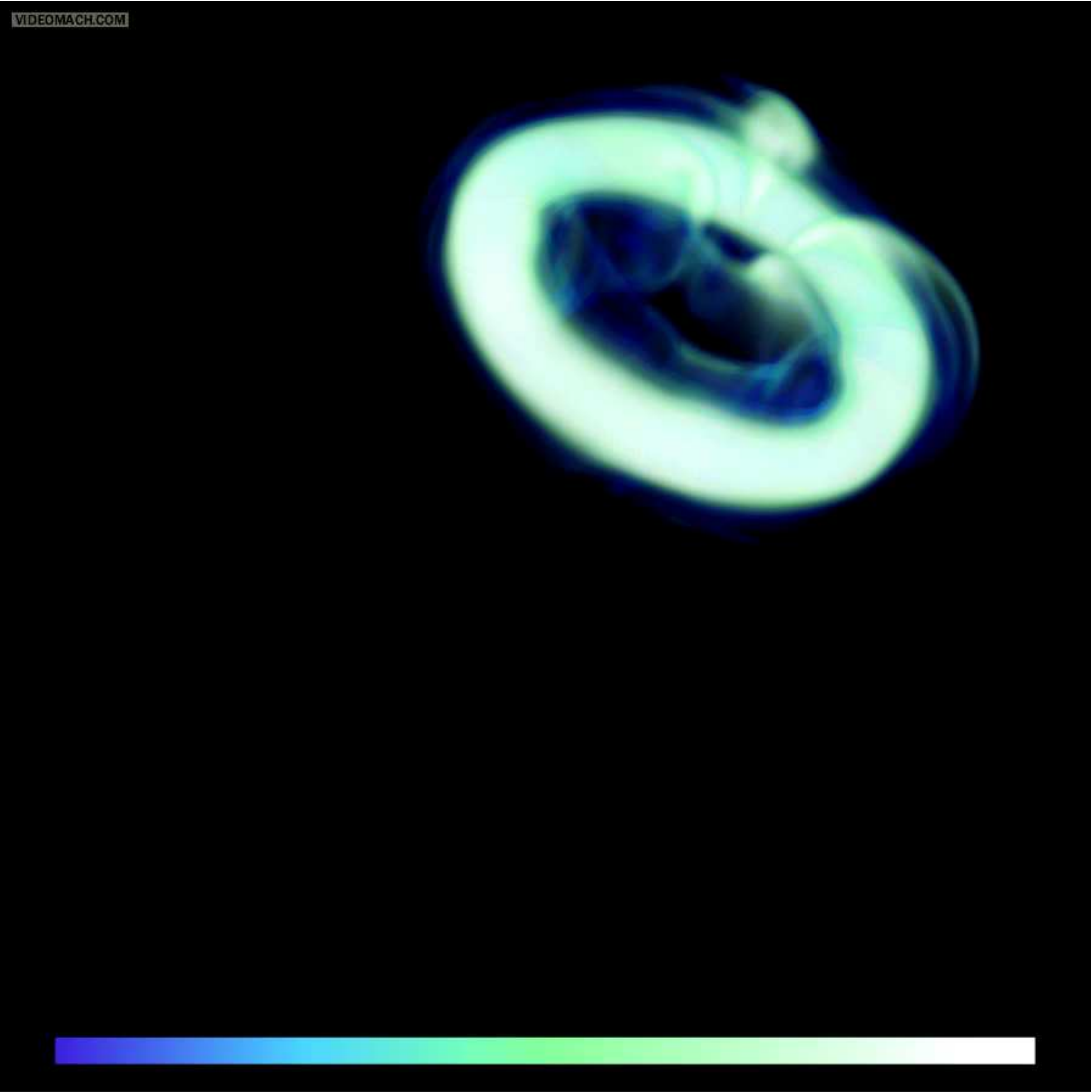}
\caption[Volume renderings of the passive ``color'' variable for 
inflated bubbles from an angle]{Same as Figure 
\ref{fig:xzinfrings}, seen from above and with the ambient 
magnetic field trending diagonally from upper left to lower right.}
%fig 5
\label{fig:angleinfrings}
\end{figure*}

\begin{figure*}
\includegraphics[width=0.5\textwidth]{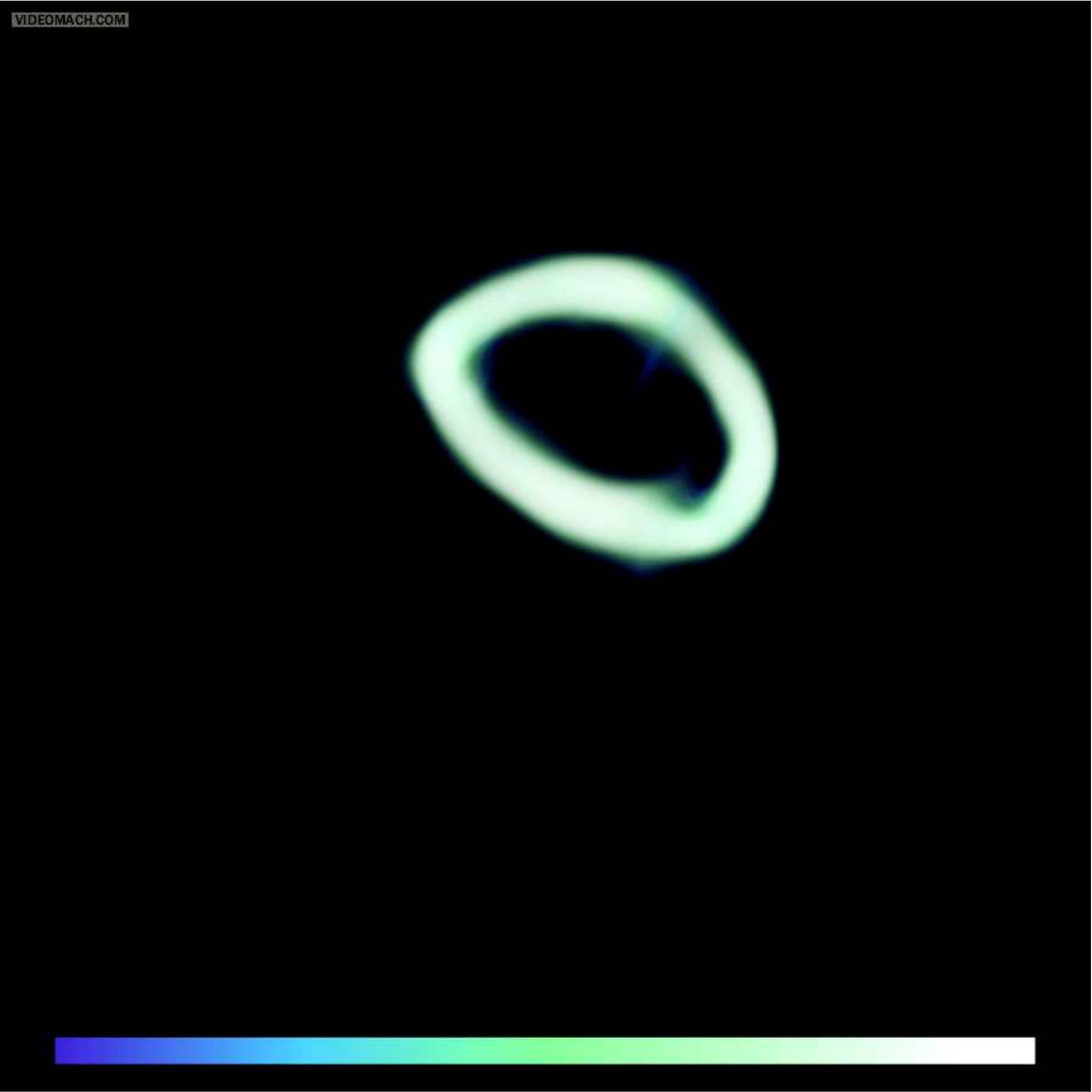}
\includegraphics[width=0.5\textwidth]{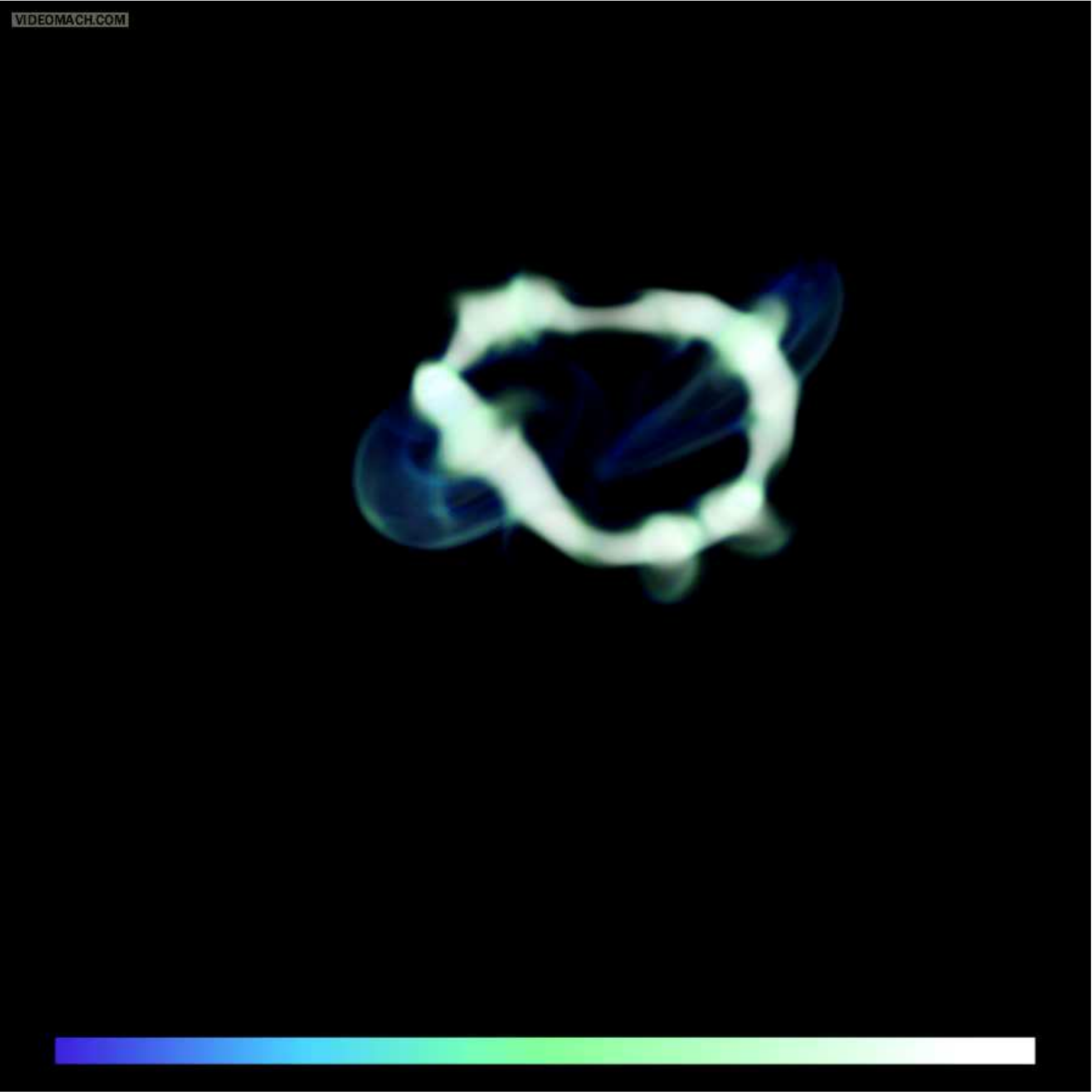}
\includegraphics[width=0.5\textwidth]{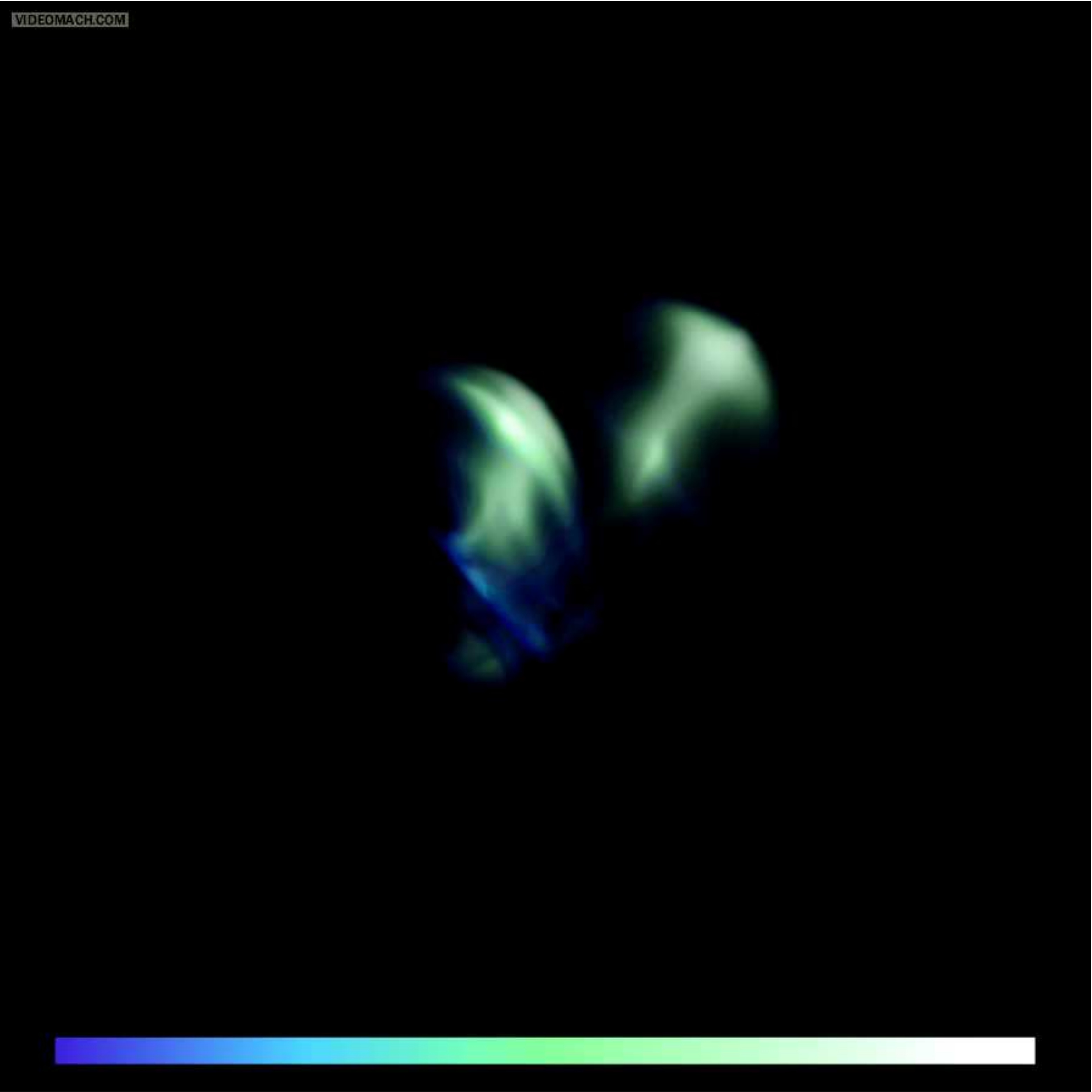}
\includegraphics[width=0.5\textwidth]{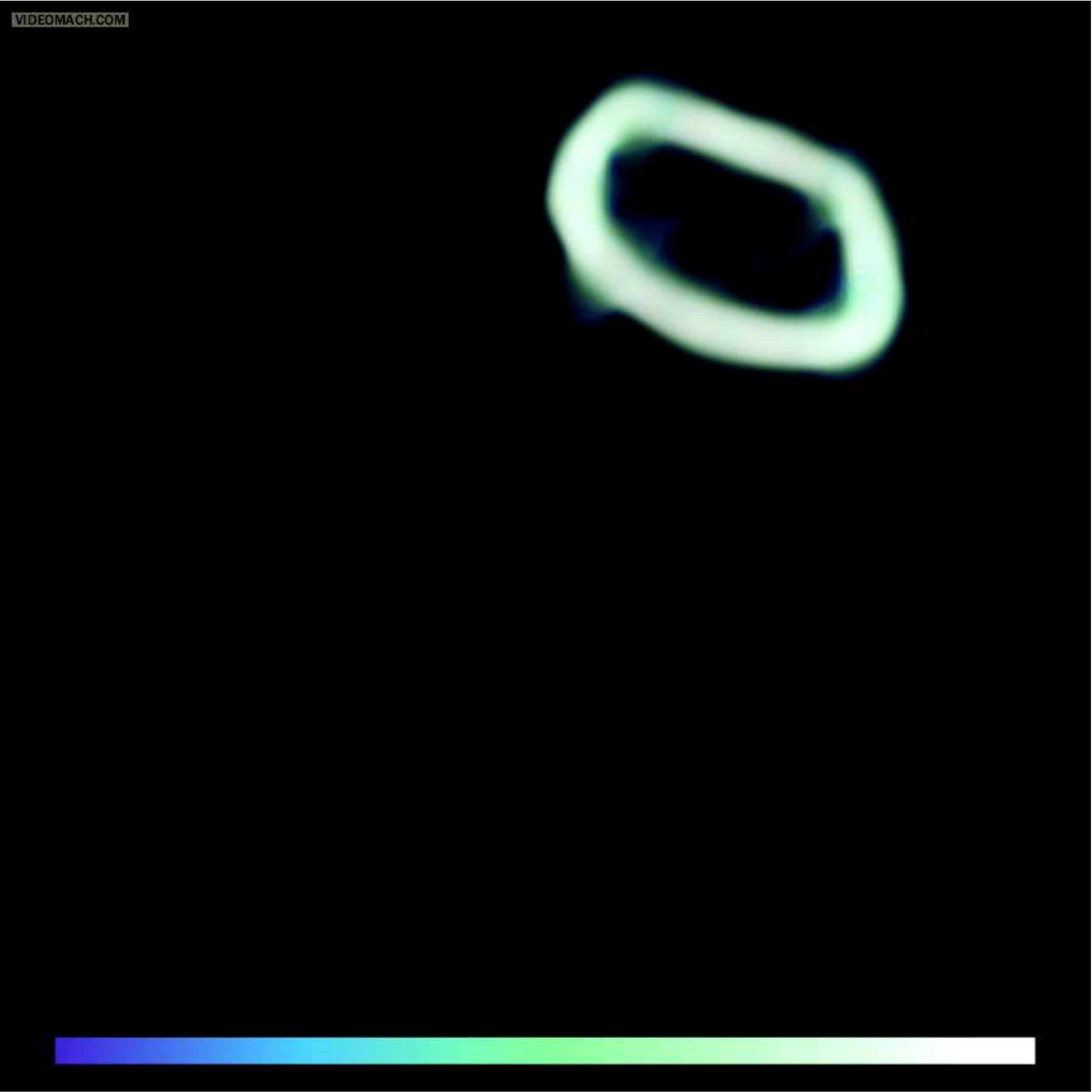}
\caption[Volume renderings of the passive ``color'' variable for non-inflated bubbles from an angle]{Same as Figure \ref{fig:angleinfrings}, except for the models without inflation. {\it Top left:} UW model. {\it Top right:} UM model. {\it Bottom left:} US model.  {\it Bottom right:} T$_B$S model.}
%fig 6
\label{fig:anglenoinfrings}
\end{figure*}

\begin{figure*}
\includegraphics[width=0.5\textwidth]{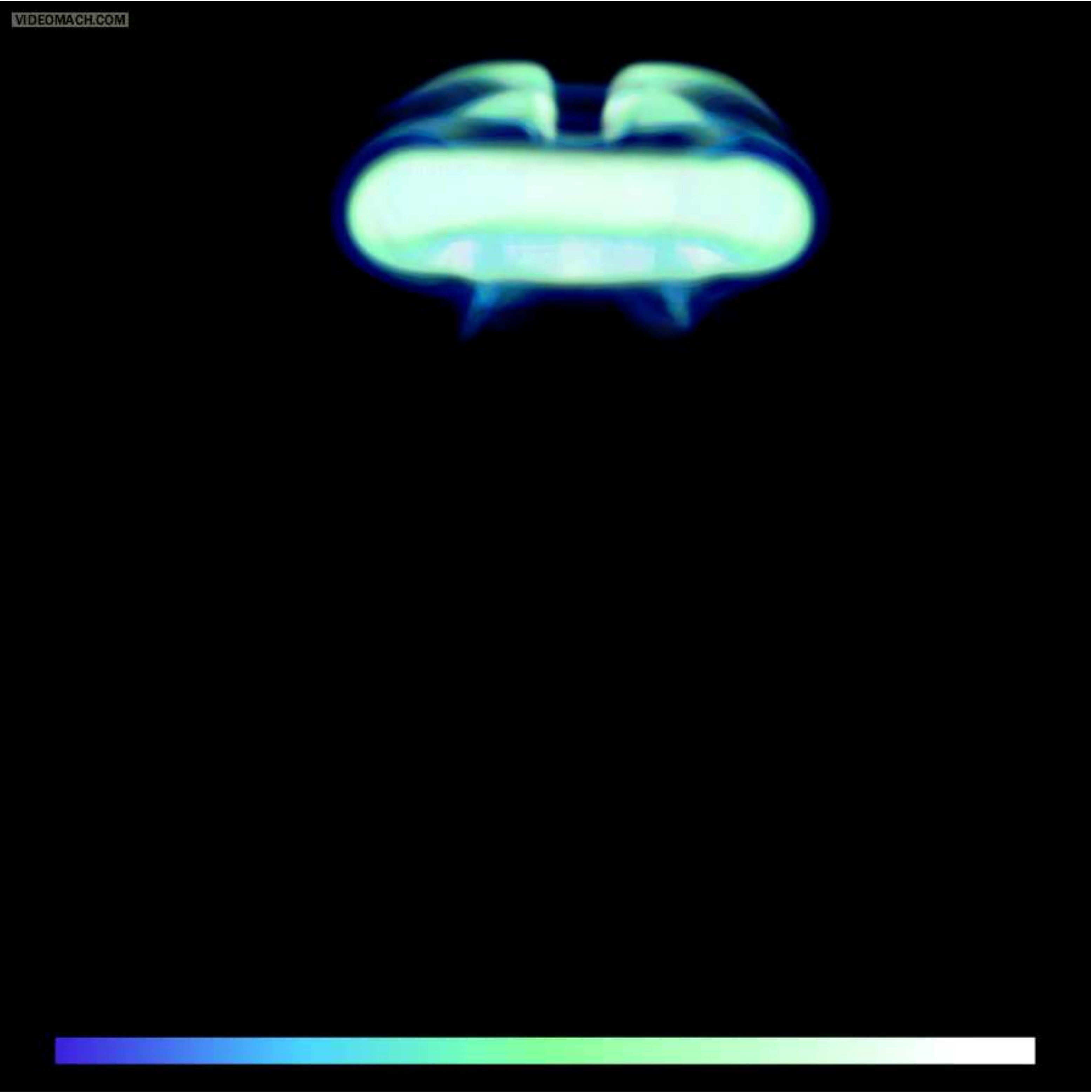}
\includegraphics[width=0.5\textwidth]{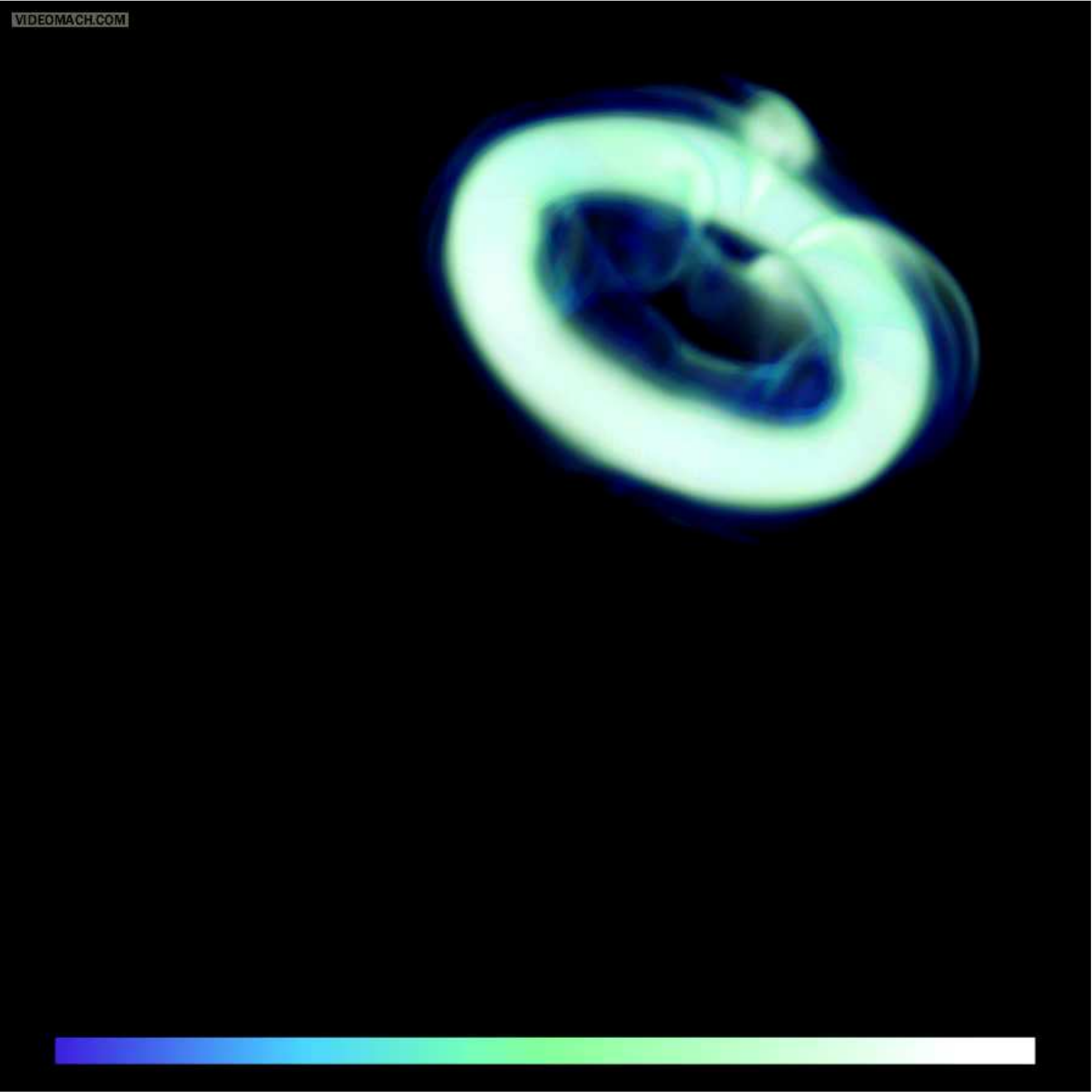}
\includegraphics[width=0.5\textwidth]{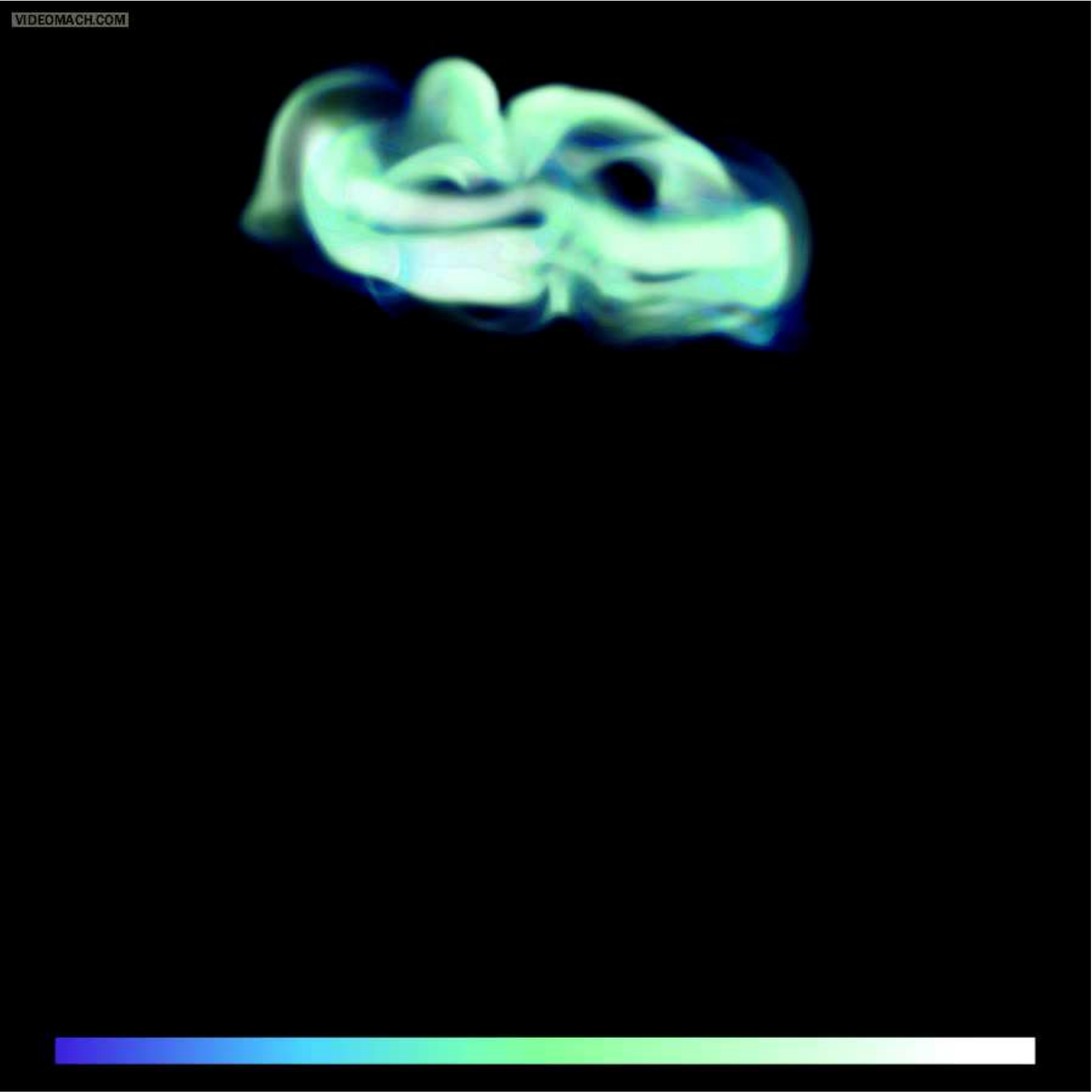}
\includegraphics[width=0.5\textwidth]{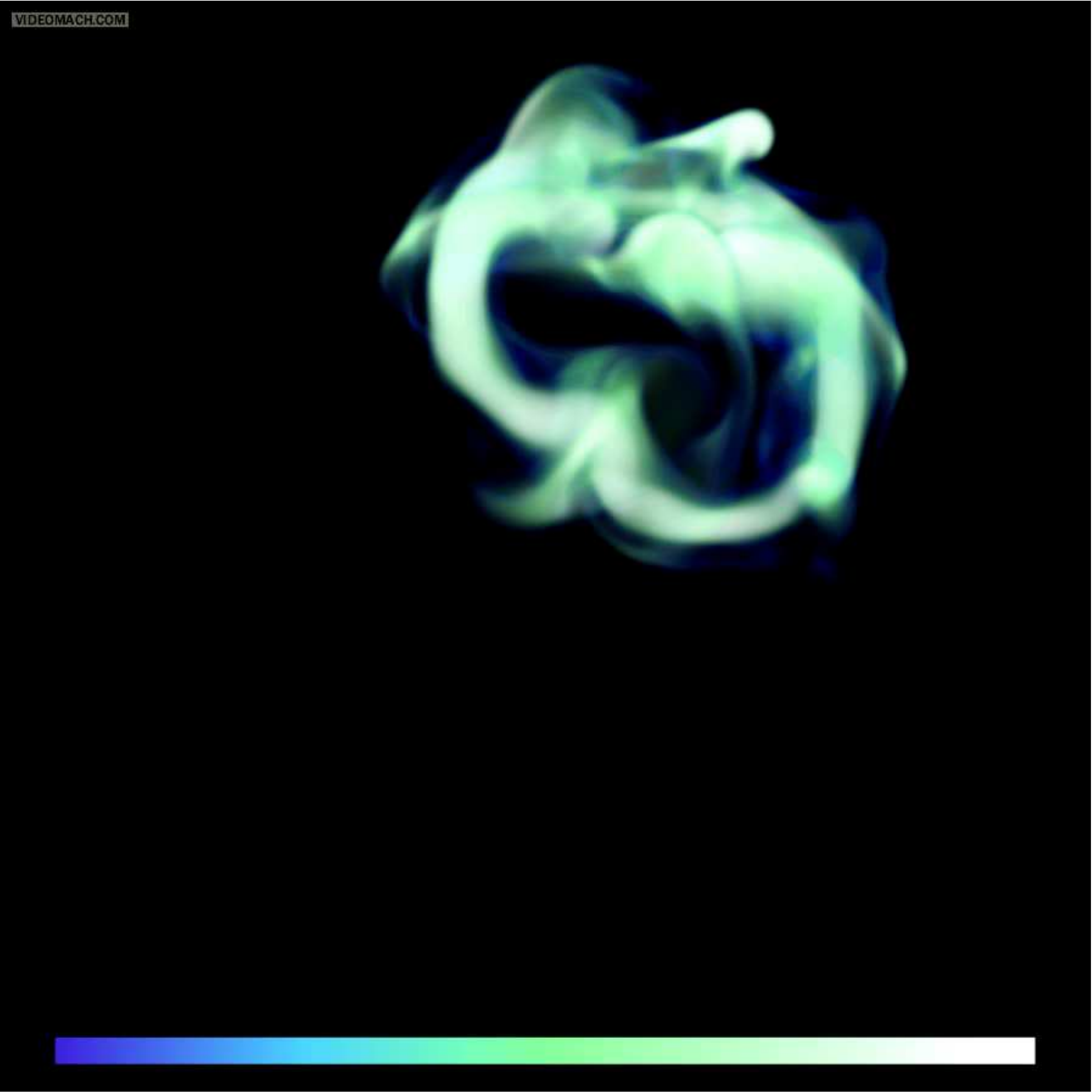}
\caption[Volume renderings of the passive ``color'' variable for inflated bubbles in tangled ambient fields]{Volume renderings of the passive ``color'' variable for inflated bubbles near the end of their evolution in a tangled ambient field.  {\it Top:} T$_B$S-I model from two different angles. {\it Bottom:} T$_L$S-I model from two different angles.  Animations of these quantities as seen from several different angles are available at http://www.astro.umn.edu/groups/compastro/ under ``Buoyant Bubbles in Galaxy Clusters.''}
% fig 7
\label{fig:tangleambientrings}
\end{figure*}

\begin{figure*}
\includegraphics[width=0.5\textwidth]{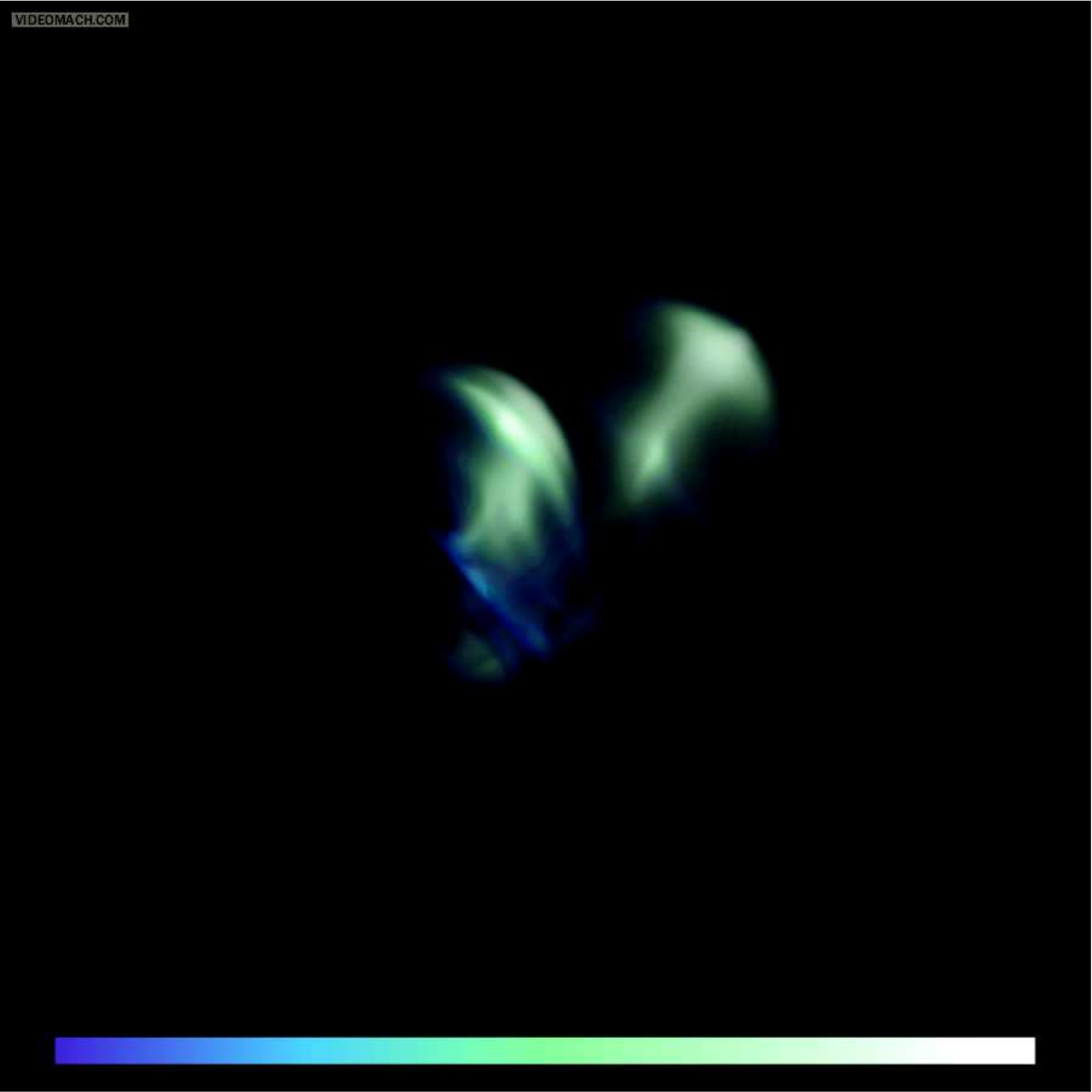}
\includegraphics[width=0.5\textwidth]{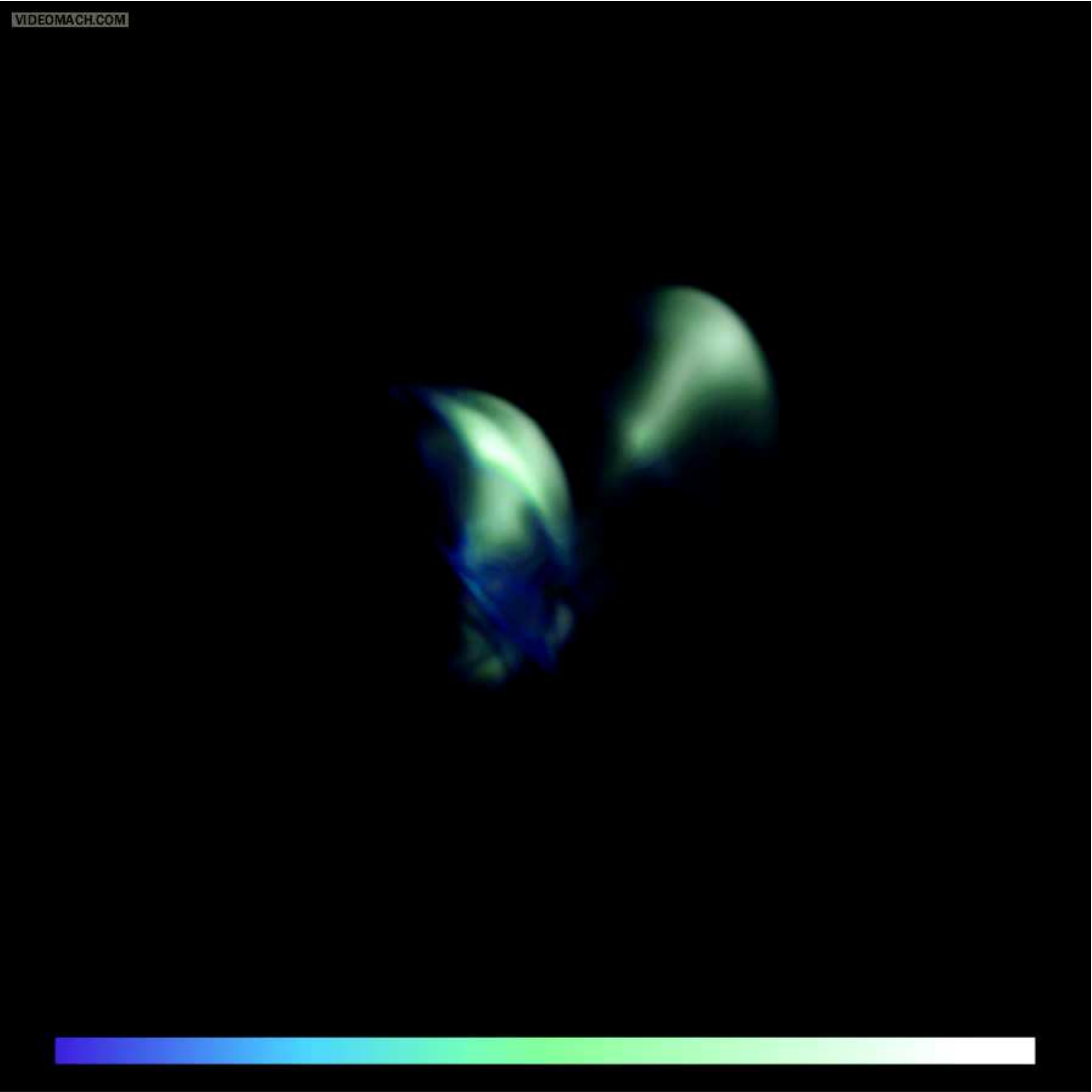}
\includegraphics[width=0.5\textwidth]{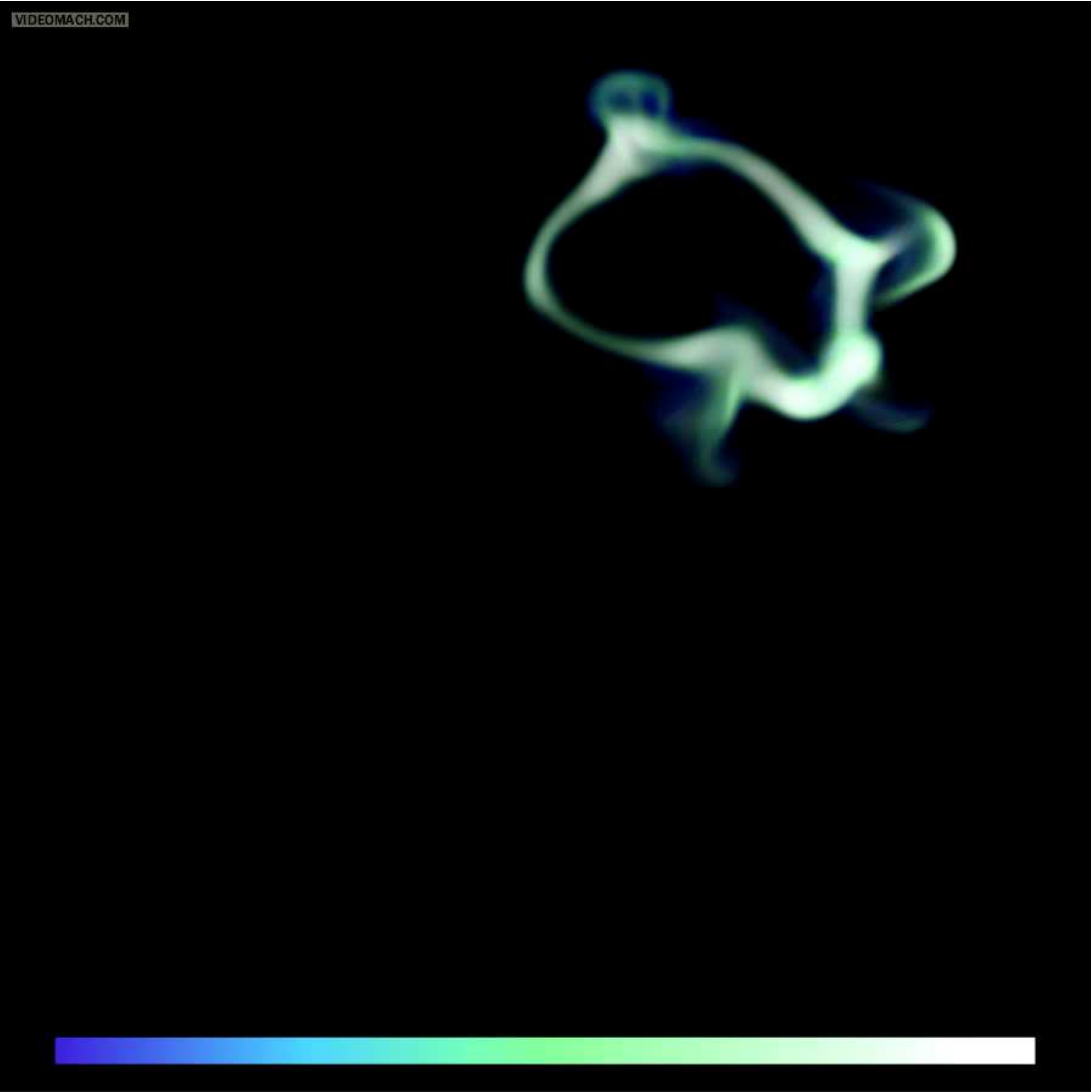}
\includegraphics[width=0.5\textwidth]{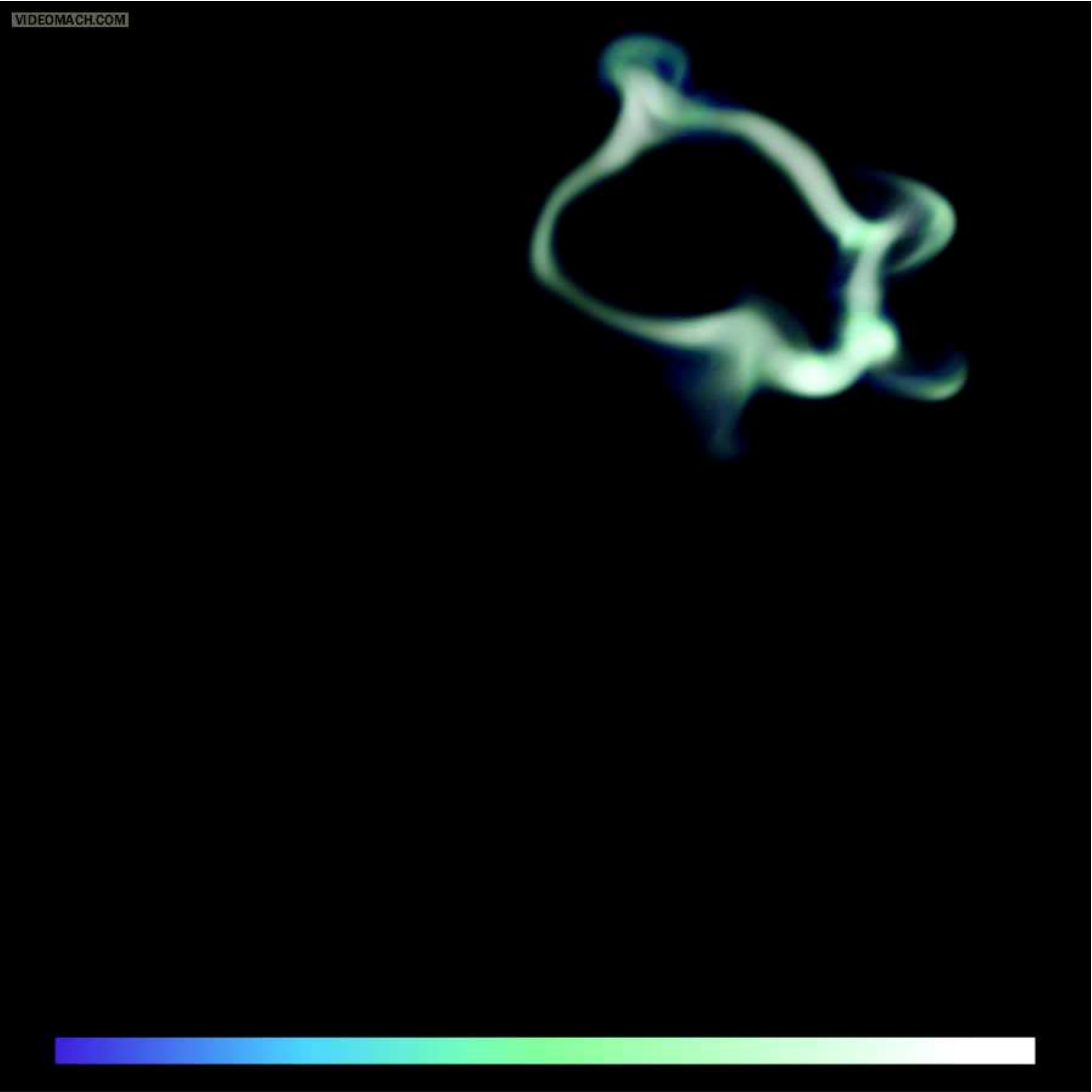}
\caption[Volume renderings of the passive ``color'' variable for bubbles with and without a tangled internal bubble field]{Volume renderings of the passive ``color'' variable for bubbles with and without a tangled internal bubble field near the end of their evolution.  {\it Top left:} US model. {\it Top right:} US-F model. {\it Bottom left:} T$_L$S model.  {\it Bottom right:} T$_L$S-F model.  Animations of these quantities as seen from several different angles are available at http://www.astro.umn.edu/groups/compastro/ under ``Buoyant Bubbles in Galaxy Clusters.''}
%fig 8
\label{fig:tangleintrings}
\end{figure*}

We conducted a series of 12 simulations, outlined in 
Table 1, designed to examine the nature of 3D bubble propagation 
and in particular to explore the effects of magnetic fields on development
of bubble shape and stability.
Figures \ref{fig:xzinfrings} - \ref{fig:tangleintrings} 
illustrate the structures of each simulated bubble at the end of its simulation.
These figures and their associated animations display volume renderings of the passive color variable $C_f$,
to highlight the volumes occupied by material that
originated within the bubble inflation region. To a good approximation this traces
the regions where the density is substantially below the neighboring 
ICM. Figures \ref{fig:xzinfrings} and \ref{fig:angleinfrings} 
show unmagnetized bubbles ``inflated'' into ICMs
spanning a range of magnetic field strengths and coherence lengths. 
Figure \ref{fig:anglenoinfrings}
shows the analogous states for ``uninflated'' bubbles created in
the same environments as those shown in Figures \ref{fig:xzinfrings} and \ref{fig:angleinfrings}. 
Figure \ref{fig:tangleambientrings} displays the forms of bubbles inflated into ICMs containing
a strong field tangled on different scales, while Figure \ref{fig:tangleintrings} shows
two examples of uninflated bubbles formed in quasi-uniform and tangled 
fields. We now outline the dominant dynamical developments
that these images illustrate and the roles of the ambient and internal
magnetic fields. Comparison of Figures \ref{fig:angleinfrings} and
\ref{fig:anglenoinfrings} shows that there are only minor differences for a given set of other parameters between inflated and uninflated bubbles for the short inflation times we used.
Primarily, the inflated bubbles are somewhat larger than the uninflated bubbles, as one would naturally expect.
The basic bubble shapes, however, are quite similar.
Specifically, ring structures (see $\S$ \ref{sec:ringformation}) are formed initially in all models, regardless of inflation, and similar bifurcation patterns (see $\S$ \ref{sec:magneticinfluences}) are seen in the US and US-I models.
Given the minor effect of a short inflation period, detailed discussion of differences between the inflated and uninflated bubbles seems unnecessary.

It is useful for this discussion to establish up front some representative timescales,
beginning with $\tau_{s,0} = h/c_{s,0}$, the characteristic sound crossing time over a
scale height in the 
ICM. This is about 15 Myr where our bubbles are initialized ($x = 13$ kpc) and increases
approximately linearly with $x$. The internal bubble sonic timescale would be 
$\tau_{s,b} = \tau_{s,0} (r_b/h) \sqrt{\eta}$, much faster than $\tau_{s,0}$.
We discuss in \S 3.3 details of the bulk upward motions of
the bubbles, where we establish their terminal upward velocity to
be typically $u_t \lesssim 1/3 c_{s,0}$. Thus, the time for a bubble to rise across
its own radius would be $\tau_b \sim r_b/u_t = \tau_{s,0} (r_b/h) (c_{s,0}/u_t)$; this is
roughly comparable to $\tau_{s,0}$ for our bubbles, which initially satisfy $r_b/h \sim 1/6$.
Timescales for possible disruption from instabilities can be estimated from the linear
growth rate of perturbations comparable in size to the bubble radius $r_b$. Using the condition
$\eta \ll 1$ the hydrodynamical R-T disruptive timescale would then be 
$\tau_{RT} \sim \sqrt{r_b/h} \tau_{s,0}$,
so $\tau_{RT}$ is somewhat
less than $\tau_{s,0}$ or $\tau_b$ for our bubbles.
The analogous disruption time for the hydrodynamical K-H instability, as derived from the dispersion relation for perturbations at a tangential discontinuity \citep[see][for example]{landaulifshitz} and letting the velocity shear be $\sim u_t$, would be $\tau_{KH} \sim (r_b/h) (c_{s,0}/u_t)(1/\sqrt{\eta}) \tau_{s,0}$.
Given $\eta \ll 1$, the K-H disruption time would be significantly longer than any of the other listed timescales.

\subsection{Bubble Morphology: Ring Formation}\label{sec:ringformation}

One striking feature of the early dynamical evolution of all the 
bubbles we simulated
is the transition from the initial spherical form into a ring or torus
around a roughly vertical axis. 
Examination of the velocity fields reveals strong flow up
through the center of the rings and a clear circulation within and
around them; i.e., the bubbles evolve into vortex rings or ``smoke rings''.
Axial symmetry is broken by the ambient magnetic field, although
the ring-like form is still evident even at late times except for the 
strong magnetic field cases.
(Note that "strong" in this case still refers to a magnetic field ($\beta \sim 100$) that is initially dynamically unimportant and comparable to the observationally inferred ICM values)
Similar evolution is evident in the 3D MHD bubble simulations
reported by \citet{ruszkowskietal07}, the 3D hydrodynamic simulations of \citet{2008MNRAS.384.1377P,2007arXiv0709.1796P}, and an axisymmetric 2D MHD bubble simulation reported by \citet{robinsonetal04}. 
Although rings were precluded in the 2D plane
symmetric bubbles of \citet{jonesdeyoung05} and \citet{robinsonetal04},
the division there of the initial cylindrical bubbles into a pair of counter rotating
line vortices with axes orthogonal to the computational plane was analogous.

Since this behavior seems to be a common element of both 
preformed and briefly inflated buoyant bubbles, it is important
to identify the cause. So far as we are aware, no physical
explanation has been offered before, although it
is straightforward.
It is easy to see, in fact, that the initial pressure balance between 
the bubble and the ICM leads directly to a strong upward push
through the bubble center. Much like a smoke ring, the return flow
around the outside of the bubble sets up the vortical flow that leads
to a relatively stable ring.
The appendix outlines two semi-analytic ways to estimate the acceleration of ICM gas into the 
bubble from below, which we have tested quantitatively against 1D simulations. 
In short, because the pressure gradient inside
a light bubble initially close to pressure equilibrium with its
environment is set by
the weight of a denser, external medium, the weight of gas inside the 
bubble itself is not sufficient to balance the pressure gradient. The light 
gas accelerates upward inside the bubble, creating a rarefaction at the lower bubble-ICM 
boundary. This propagates across the bubble in a time
$\tau_{s,b} \ll \tau_b$.

In response to the lowered pressure inside the
bubble rarefaction, the lower bubble 
boundary, or contact discontinuity (CD), accelerates 
upward at a rate approximately given by (see equations \ref{cdone} and \ref{cdtwo})
\begin{equation}
\ddot{x}_{CD} = -g(x)\sqrt{\rho_0(x)/\rho_b} \sim -g/\sqrt{\eta},
\label{bubbacc}
\end{equation}
where $g(x)$ is the local gravity and, again, $\eta = \rho_b/\rho_0(x_b)$. 
Representative timescales for CD motion 
suggested by equation \ref{bubbacc} are
$\tau_{CD} = \sqrt{h/|g|}~\eta^{\frac{1}{4}} = (h/c_{s,0})~\eta^{\frac{1}{4}} = \eta^{\frac{1}{4}}~\tau_{s,0}$, and $\tau_{CD,b} = \sqrt{r_b/|g|}\eta^{\frac{1}{4}} = \sqrt{r_b/h} \tau_{CD}$. 
These measure how long it takes the bottom of the bubble to rise up through a
scale height and across the scale of the bubble, respectively.
Note that so long as $\eta \ll 1$, $\tau_{CD,b} < \tau_{RT}$, and 
this development within the bubble will be faster than disruptive 
R-T instabilities.
If $\tau_{CD,b} < \tau_b$, there is the potential for the bottom of the bubble
to catch up to the top while the bubble is still close to its initial 
location, which is what we observe to happen in our simulations.

For the bubble actually to develop into a vortex ring 
the center of the bubble bottom must move upwards faster than the
bottom regions near the bubble perimeter; that is, the bottom CD
must transform from its initial concave upward shape into a convex shape.
In a hydrostatic ICM 
that outcome results in equation \ref{bubbacc} from the fact that 
$|g|/ \sqrt{\eta}$ is generally greatest at the lowest point along the CD.
The relative difference in $\ddot{x}_{CD}$ grows with $r_b/h$,
so vortex rings form more easily when the bubbles are
larger. In our simulations, inflated bubbles are larger
compared to uninflated bubbles when they are ``released'',
so they are somewhat more strongly driven into the ring transformation.
If we ask that the lowest point of 
the bubble CD is driven up to overtake the side of the bubble within a time $\tau_b$, using
equation \ref{bubbacc} and assuming the acceleration is constant,
we can roughly estimate the condition for ring formation to be
$r_b/h \gtrsim \eta^{\frac{1}{4}} (u_t/c_{s,0})$, which is only weakly
dependent on $\eta$. For our models, this constraint would be $r_b/h \gtrsim 1/9$.
This condition is satisfied by our bubbles and also 
by the bubbles simulated by \citet{robinsonetal04} and \citet{ruszkowskietal07}.

As a concrete illustration of this effect we can construct a simple 
2D planar model assuming
a bubble of initial uniform density, $\rho_B$, in an  isothermal
ICM with constant gravity in the $-x$ direction, so that 
$\rho_0(x) = \rho_0 \exp(-x/\tilde h)$ and $P_0(x) = P_0 \exp(-x/\tilde h)$,
with $\tilde h = c_{s,0}^2/(\gamma g) = h/\gamma$. The $\gamma$ factor appears 
because we defined the scale height, $h$, using the adiabatic sound speed.
The gravity in our simulations is not constant, but scales approximately
as $g \propto 1/x$, as noted above. This actually slightly
enhances the effects described in this simpler model. We were unable to find a closed
solution for the variable gravity case, but verified numerically
that the CD acceleration is similar to the constant gravity model outlined here.
For our simple analytic model we parametrize the bubble density to be $\rho_B = \eta \rho_0$,
and describe the initial shape of the bottom CD by the curve, $s(y)$. 
It is also convenient to define time scale for each point along the CD,
\begin{equation}
\begin{aligned}
\tau(s) &= \frac{\tilde h~\eta^{\frac{1}{4}}\sqrt{\gamma} }{c_{s,0}}~ e^{\frac{s(y)}{4 \tilde h}}\\
 &= \frac{\tau_{CD}}{\sqrt{\gamma}}e^{\frac{s(y)}{4 \tilde h}}.
\end{aligned} 
\label{eq:tau}
\end{equation}
Note that $\tau(s)$ increases strongly as $s(y)$ increases upwards.
Assuming equation \ref{bubbacc} applies throughout the CD
evolution, we obtain an estimate for the position, velocity, and acceleration of the CD to time t to be
\begin{equation}
\begin{aligned}
x(y) &= 2 \tilde h \ln \left[\frac{(e^{ t/\tau} + 1)^2}{4e^{ t/\tau}}\right] + s(y)\\
\dot{x}(y) & = 2 \frac{\tilde h}{\tau} \left[ \frac{e^{t/\tau} - 1}{e^{t/\tau} + 1}\right]\\ 
\ddot{x}(y) &= 4 \frac{\tilde h}{\tau^2} \left[ \frac{e^{ t/\tau}}{(e^{t/\tau} + 1)^2} \right].
\end{aligned}
\label{eq:cdevolv}
\end{equation}
At the start, when $t = 0$ and $x(y) = s(y)$, the acceleration is 
$\ddot{x}(y) = (|g|/\sqrt{\eta}) exp(-s(y)/(2 \tilde h))$.
By the time $t = \tau(s)$, the point 
initially at $s$ has risen about half the scale height, $h$.
If the bubble is initially spherical this timescale is shorter
by a factor $exp(-r_b/(4\tilde h))$ at the lowest point on the bubble than 
at the elevation of its center.

\begin{figure*}[t]
\includegraphics[width=\textwidth]{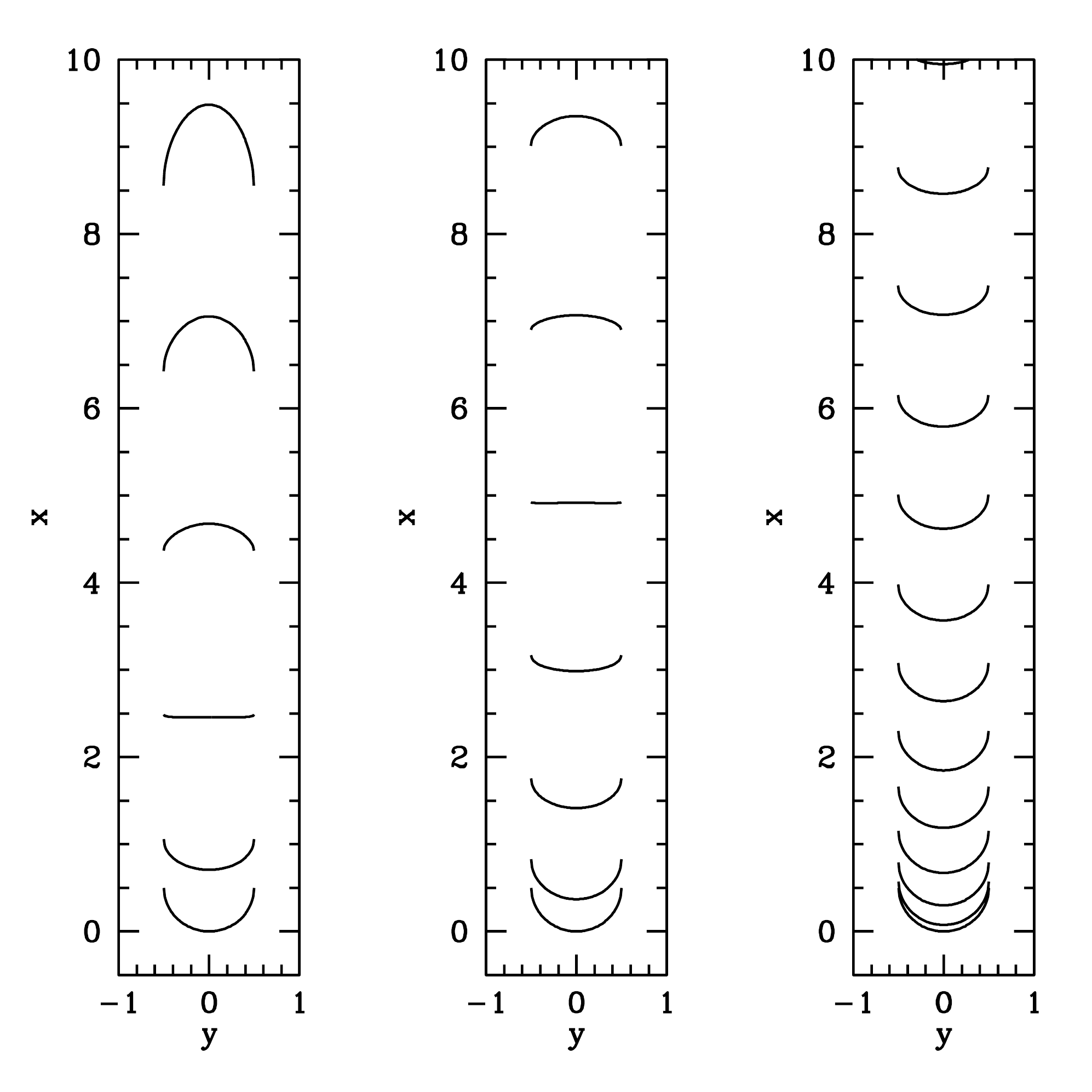}
\caption[Semi-analytic model for the evolution of the lower surface of 
a buoyant bubble for varying ICM scale heights]{Semi-analytic model 
for the evolution of the lower surface of a buoyant bubble for varying 
ICM scale heights. {\it Left:} $h=2r_b$. {\it Center:} $h=4r_b$. 
{\it Right:} $h=20r_b$.  The two axes are 2D spatial dimensions in 
units of the bubble diameter, where $+x$ is the direction of bubble 
propagation.  Time increases from bottom to top, with a time interval of $\Delta t \approx 0.08 \tau_{CD}$ between consecutive CD shapes.}
%fig 9
\label{fig:bubbleprof}
\end{figure*}

We illustrate this behavior in Figure \ref{fig:bubbleprof},
where the evolution described by equation \ref{eq:cdevolv}
is shown for three initially semi-circular, concave
upward CDs ($s(y) = x_b - \sqrt{r_b^2 - y^2}$).
The three cases vary the bubble radius to scale height
ratio; namely, left to right
$h/r_b = 2$, $h/r_b = 4$, and $h/r_b = 20$. The middle case
is closest to our simulated bubbles. 
Each case shows multiple CD shapes, corresponding to different times in the CD evolution with a time interval of $\Delta t \approx 0.08 \tau_{CD}$ between consecutive shapes. 
Except for the third case,
where the scale height is much larger than the size of the bubble,
the CD is quickly inverted to a convex upward shape, consistent with
what is seen in the simulated bubbles. 
While this model is a simplification of the real dynamics in 2D or 3D, it
does demonstrate the basic initial physical behavior of the bottom of the
bubbles. 

As a result of this behavior the ambient gas drives upwards in the
center of the bubble, while the buoyant rising of the bubble body
establishes a downward flow around the outside, thus generating
a vortex ring. Our previous comparison with the
growth rate for disruptive R-T instabilities showed that this ring
development is faster when the bubble density contrast
is large.
We expect, then, any light, initially quasi-spherical bubble with
$h \la r_b$ to be dominated by this effect, so long as the bubble
inflation process 
is terminated before the bubble top has risen through a scale height,
or roughly an interval several times $\tau_{s,0} = h/c_{s,0}$.
As mentioned previously, when the inflation process is maintained over intervals longer
than this a de Laval nozzle forms and bubble material expands
into the upper atmosphere at speeds 
$\sim c_{s,b} \sim c_{s,0}/\sqrt{\eta}$ \citep{jonesdeyoung05}.
This flow is too fast to be overtaken
by the lower boundary moving upwards at a speed slower by a factor
$\sim \eta^{\frac{1}{4}}$ according to equation \ref{eq:cdevolv}.
If the bubble inflation takes place at the center of the ICM, 
if the initial bubble was large compared to the local scale height, or if it
involved a driven flow by a jet from outside the bubble, then this dynamical
behavior would not be relevant \citep[\eg][]{reynoldsetal02, bassonalexander03, 
ommaetal04,ommabinney04,oneilletalinprep}.
On the other hand, \citet{2008MNRAS.384.1377P, 2007arXiv0709.1796P} have noted that radially-directed Kutta-Zhukovsky forces \citep[see][for example, for a discussion of Zhukovsky's theorem]{landaulifshitz} may characteristically transform flattened bubbles into vortex rings regardless of the details of inflation.
This is distinct from the process we have identified, however, in that Kutta-Zhukovsky forces require the presence of both circulation and bubble asymmetry to produce a vortex ring.
In contrast, the mechanism of ring formation that we describe operates immediately, even on initially spherical and motionless bubbles, so long as their pressure profiles are set by the equilibrium structure of the comparatively dense ICM.

We close this portion of the discussion with two simple
observations about rings formed in unmagnetized media. The first is that, in our models the rings in the UW and UW-I simulations
had roughly steady tube radii, with $r \sim r_b$.
The tube radius was larger in the UW-I case, basically because
it was bigger when ``released''.
Second, in the absence of magnetic fields the ring-like bubble structures
formed in our simulations were relatively stable over the simulation
times, although they developed some bending mode oscillations, probably from the Widnall
instability \citep{widnall73,widnall74} that is known to affect vortex rings, and surface irregularities likely due to K-H instabilities. 
The observed coherent behavior is expected
for vortex rings when the effective Reynolds number, $R_e$ is less
than about $10^3$, although for much higher Reynolds number flows vortex
rings are seen experimentally to be unstable to turbulent break-up 
\citep[\eg][]{max72}. It is not possible to define a
rigorous Reynolds number for calculations with our inviscid MHD code;
however, based on code tests and the fact that
the tubes were spanned in our simulations by $\sim 30$ cells, we
can estimate an effective (fast mode) Reynolds number for flow
around the ring to be $R_e \sim 500-1000$ \citep{ryu00}.
The 3D simulations by \citet{ruszkowskietal07} applied an explicit
viscosity, with a Reynolds number, $R_e \sim 2000$. The ring structure
seen in their unmagnetized bubble simulation appears to have survived
intact for $\approx 100$ Myr, although those authors also note the
presence of small scale
K-H instabilities. Thus, if such rings are indeed
destructively unstable on timescales of interest, substantially higher 
resolution than any currently
available seems to be necessary to capture that behavior.

\subsection{Bubble Morphology: Magnetic Influences}\label{sec:magneticinfluences}

The early evolution of the bubbles (\ie ring formation) was essentially independent of
the magnetic field properties we applied. That is expected, since
the initial Maxwell stresses from even the strongest ICM field, 
with $\beta = 120$, were insignificant. That result
mirrors our 2D simulations as well.
We note, however, that \citet{ruszkowskietal07} found varying degrees of ring formation between their magnetized and non-magnetic models.
While it is clear that all of their models, like ours, are subject to the initial forces that would cause ring formation, their magnetized models featured initial fields (${\bar \beta} \sim 40$), both internal and external, that were apparently sufficiently strong to prevent the completion of this process.
Although we have not attempted here to model fields strong enough to disrupt ring formation, our simulations do show that the inclusion of even a moderate ambient magnetic field ($\beta = 3000$) can have a clear impact on evolution of the ring structures, provided the coherence scale for the field exceeds the size of the bubble. 

\begin{figure*}[t]
\includegraphics[width=\textwidth]{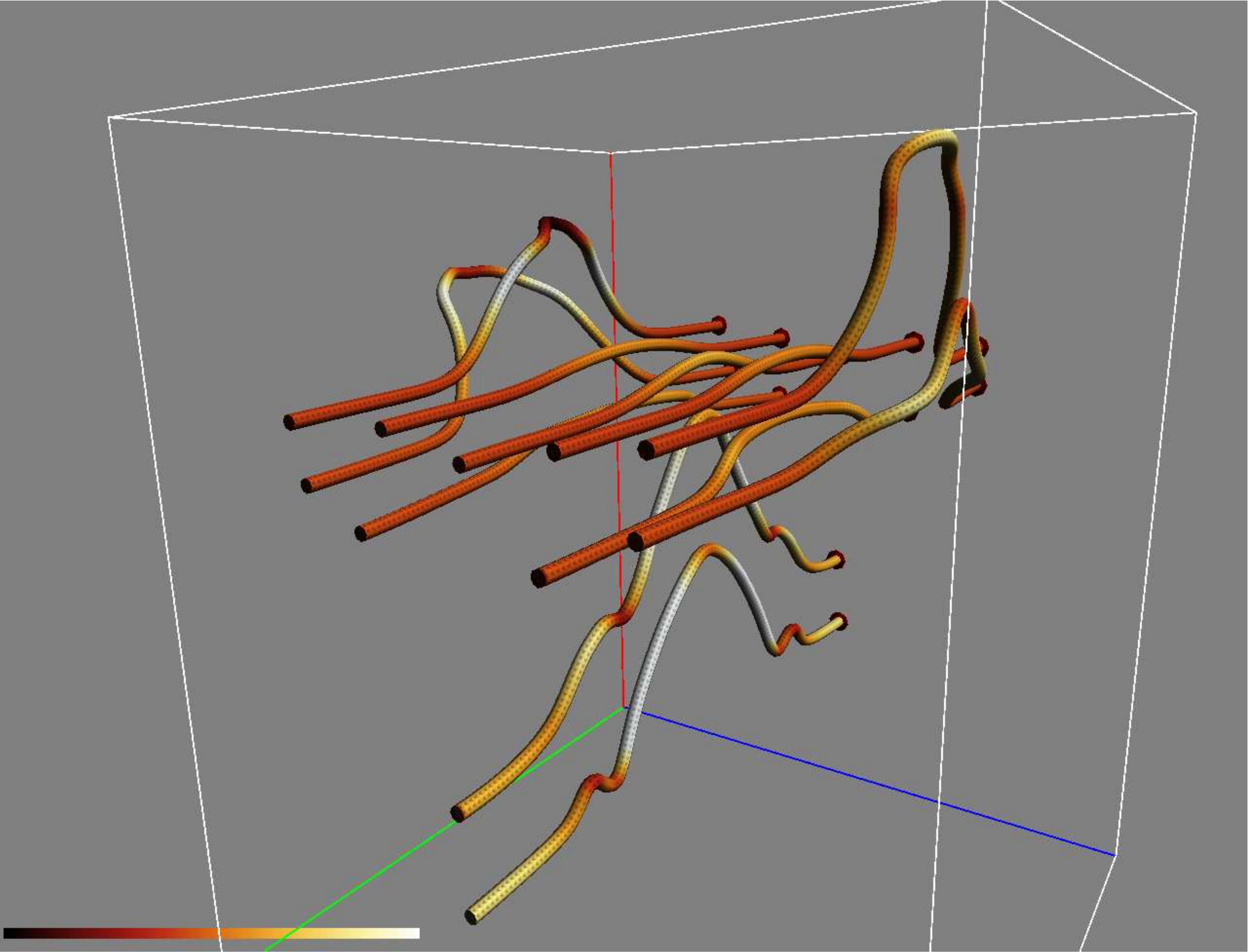}
\caption[Field lines in the US-I model.]{Field lines in the US-I model near the end of the simulation time, showing the magnetic field structures that have bifurcated the bubble.  The color on the lines represents field strength (lighter=stronger).}
%fig 10
\label{fig:b120fv}
\end{figure*}

\begin{figure*}[t]
\begin{center}
\includegraphics[width=0.7\textwidth]{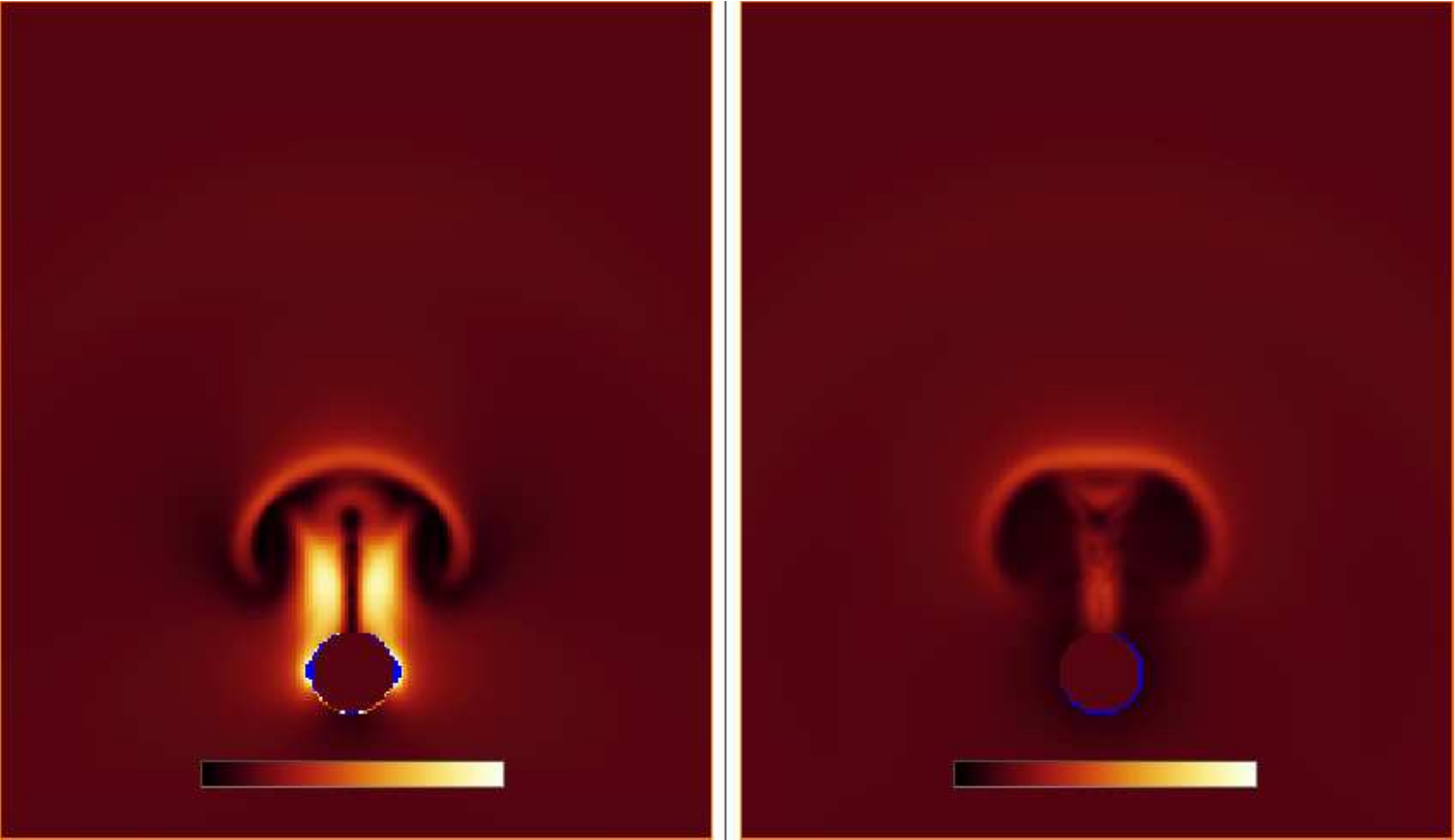}
\includegraphics[width=0.7\textwidth]{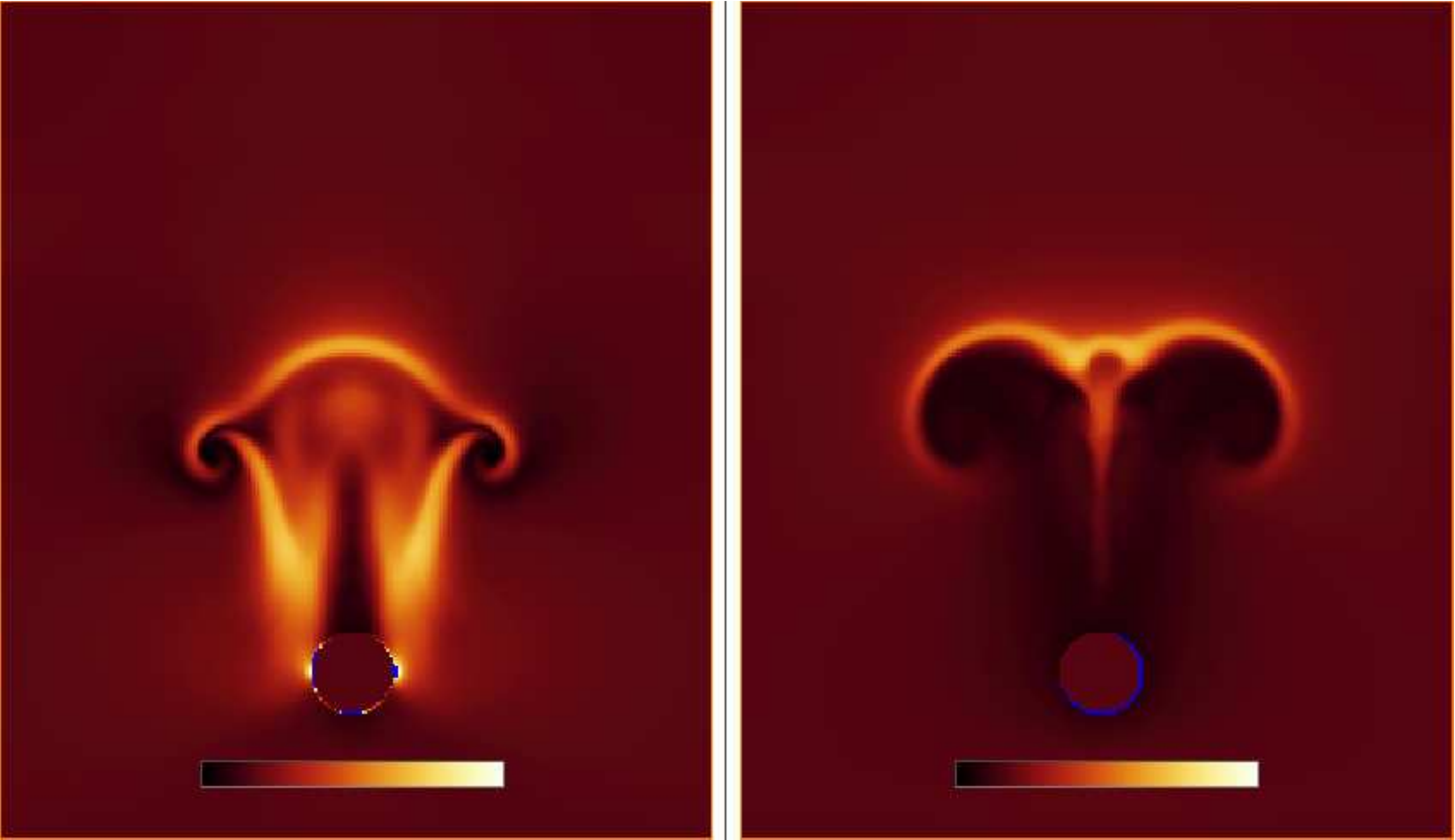}
\includegraphics[width=0.7\textwidth]{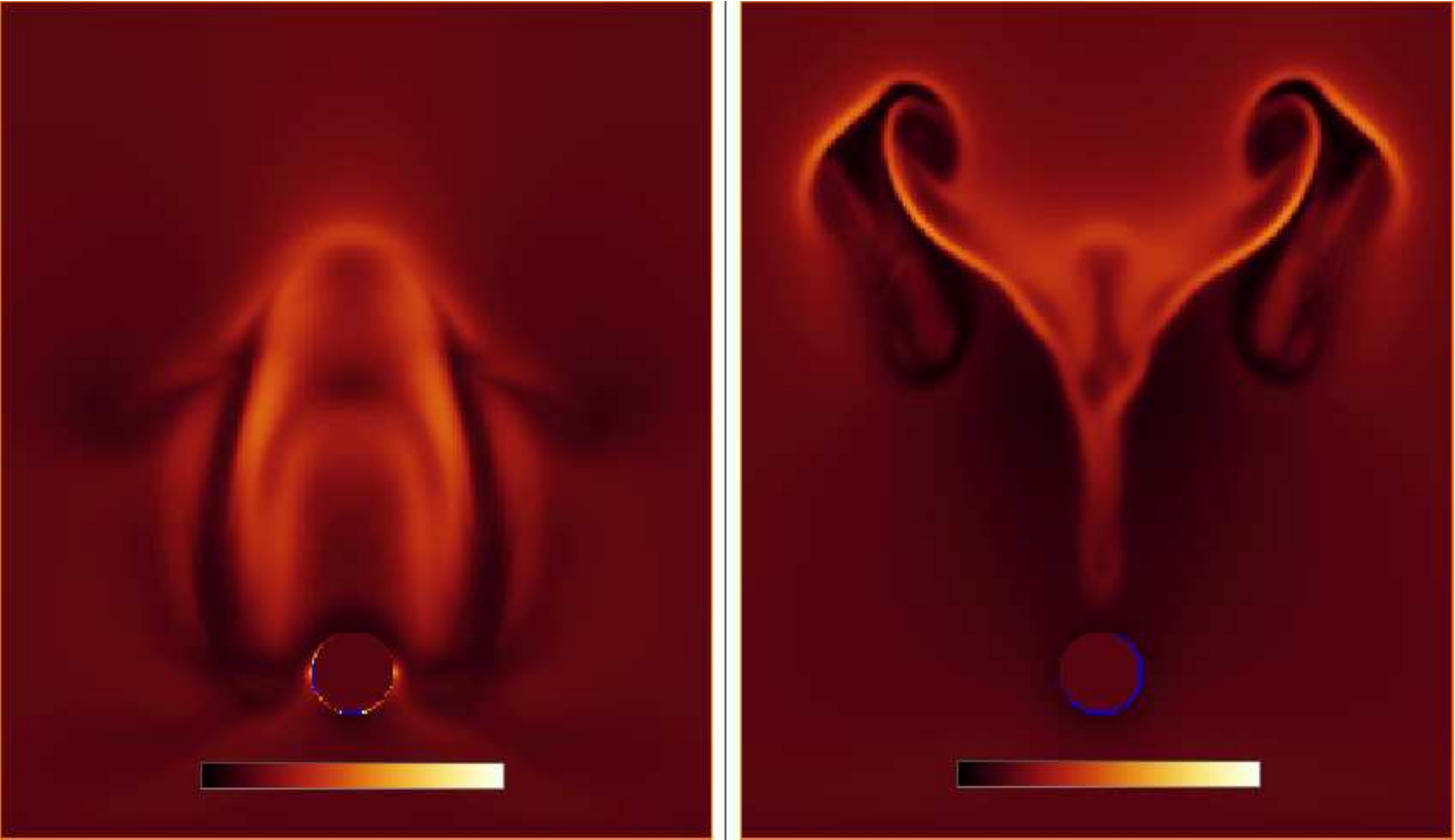}
\caption[Midplane slices showing the ratio of the magnetic field strength to its initial value ($B/B_0$) for the US-I model]{Midplane slices showing the ratio of the magnetic field strength to its initial value ($B/B_0$) for the US-I model in the $x-y$ ({\it left column}) and $x-z$ ({\it right column}) planes.  The top row shows $t=25~$Myr, middle $t=50~$Myr, bottom $t=100~$Myr.  The colorbar ranges from [0:6].}
%fig 11
\label{fig:amirab120}
\end{center}
\end{figure*}

Consider first the
simpler cases with quasi-uniform ambient fields. The ``strong'' field
cases, US, US-I, with $\beta = 120$ show this behavior most 
obviously, so we examine those first.
The lower-left images in Figures \ref{fig:xzinfrings} and \ref{fig:angleinfrings} and their associated animations 
illustrate that these bubbles developed a wishbone shape; that is, the bubbles were bifurcated.
The initial ring became elongated in the horizontal direction orthogonal to the ICM field. 
As the animations show, the upward motion of the central portion of the ring aligned with the ICM field was stalled and then reversed, while two plumes of bubble material on opposite sides of this bifurcation continued to rise.
This behavior results when the external field gets trapped in R-T or the aforementioned Widnall instability-induced oscillations of the upper boundary of the bubble.
Initial upward motions stretch and amplify those field lines, which are anchored in the ICM away from the bubble, until the magnetic tension is sufficient to resist continued upward motions.
Eventually, this magnetic tension exerts a force that exceeds the buoyant force, causing those portions of the bubble in which the ICM field is embedded to reverse direction and sink.
R-T and K-H instabilities in the plane of the field and the bubble motion (the $x-y$ plane) become inhibited, of course, even though perturbations orthogonal to the field are not. 
We verified that in these regions $\beta$ can approach unity and, more importantly, that the Maxwell stresses (\ie $j\times B$) are comparable to and sometimes dominate the gas pressure and gravitational forces ($\nabla P$ and $\rho g$) there during this period.
This highlights the fact that the $\beta$ parameter is not sufficient to evaluate the
role of the magnetic field, since the relative stresses depend
not only on the magnitude of the magnetic field, but also
on the scale and direction of its variation.
On the other hand, since bubble motions in the $x-z$ plane are not magnetically
inhibited, portions of the bubbles slide through the field lines,
creating the plumes.
The magnetic field lines and intensities during this stage of the
evolution of model US-I are illustrated in Figures \ref{fig:b120fv} and \ref{fig:amirab120} respectively.

This evolution is qualitatively similar to that seen for dense supersonic, spherical clumps modeled by \citet{gregori99}, where an initially uniform ISM magnetic field wrapped around the clump, eventually also bifurcating it.
Like our 3D bubbles, the boundaries of the  \citet{gregori99} clumps were relatively sharp.
On the other hand, the outcome of \citet{gregori99} appears different in some ways from the simulation results reported by \citet{dursi08} of dense, centrally condensed, spherical clumps moving from an unmagnetized environment into a medium containing a laminar ``draping'' magnetic field.
The latter computations were designed to study formation of cold fronts during cluster mergers. 
Those clumps developed a strongly magnetized layer on their noses, but at least on timescales required for them to sweep by their own masses they were not bifurcated. 

Bifurcation behavior seen in the US and US-I simulations is also visible in the UM and UM-I models, although it is much less pronounced, due to the much weaker initial fields.
In the upper-right images in Figures \ref{fig:xzinfrings} and \ref{fig:angleinfrings} and their associated animations, we can see that vortex shedding leads to a buildup of material in the direction orthogonal to the uniform field.
This asymmetry appears to be caused by the same variety of magnetic stresses that are present in the US and US-I simulations, and over longer times it is likely that these rings would also become bifurcated.
It is also clear that the portion of the UM bubble oriented along the field lines is less elevated than the rest of the structure, which is also similar to the behavior of the bifurcated bubbles.

As emphasized by \citet{ruszkowskietal07} the influence of the magnetic 
field on bubble evolution depends also on the geometry of
the field. In particular, if the ambient magnetic field is tangled on 
scales less than or comparable to the scale of the bubble, then field 
lines
are not effectively stretched and amplified as the bubble
lifts upwards, in comparison to what happens if the field
structure is laminar. We explored that issue further in our
experiments using our
simpler tangled strong field models, T$_B$S and T$_L$S outlined in \S 2.1.
In short, these simulations have ICM field strengths comparable to
the US models, but field lines that are tangled on scales 
of the initial bubble circumference and a factor of five
greater, respectively.
Figure \ref{fig:tangleambientrings} shows a comparison of the 
bubble morphologies for the T$_L$S-I and T$_B$S-I models,
while comparisons with uniform field cases are available
in Figures \ref{fig:xzinfrings}, \ref{fig:angleinfrings},
\ref{fig:anglenoinfrings}, and \ref{fig:tangleintrings}.

In both the T$_B$S-I and T$_L$S-I cases the ring structures
that form remain intact to the end of the simulation, whereas for
the analogous US-I simulation the ring was bifurcated, as discussed
previously. For the smaller scale field structure (T$_B$S-I), 
in fact, the bubble ring is hardly different from the unmagnetized
case. An examination of the magnetic field properties
explains this simply, since it turns out the field is generally
not amplified compared to initial conditions. In fact, as a result of
reconnection that takes place in the flow around the bubble, the
$\beta$ of the ICM plasma near the bubble is actually
increased.

The situation is more complex in the  T$_L$S-I simulation. While
the ring structure survives to the end of the calculation,
there are localized regions where the field is strongly amplified
by stretching in response to the bubble's rise. This leads to
substantial distortions in the bubble, but in the absence of coherence
in the field over very large scales the dynamical impact of the
Maxwell stresses is irregular. It is likely, however, that over
still longer timescales this bubble would be disrupted by
the magnetic field. Our results on the effects of
field tangling are, therefore, in qualitative
agreement with those of \citet{ruszkowskietal07}.

Having noted that magnetic reconnection can increase the value of $\beta$ by reducing field strengths near the bubble in our models of tangled ICM fields, it is worth briefly detailing the nature of this process in our simulations.
At the most basic level reconnection is a topological transformation of the magnetic field made possible by local dissipation where adjacent fields reverse or braid \citep[\eg][]{laz08}, setting up concentrated currents.
Like other general purpose MHD codes our numerical algorithms include no explicit prescription for reconnection, per se; rather, reconnection is enabled by the effects of numerical dissipation where fields reverse or braid close to the resolution scale of the computational grid.
Although numerical in nature, this process includes essential physics involved in reconnection; namely, it exactly conserves energy and momentum, and maintains a divergence free field, while allowing such physical, reconnection-related, nonideal MHD processes as tearing modes to develop.
Consequently, the code includes the effects of physical reconnection by canceling field and channeling field energy and stresses into thermal and kinetic energy and momentum appropriate to the details of the flow and the field.
In our models of uniform ICM fields, we expect reconnection to play no significant role since localized field reversals are rare (see the field configuration in Figure
\ref{fig:b120fv}, for example).
In our models of tangled ICM fields, however, reconnection surely does take place, since local reversals and braiding are common. 
Reconnection, therefore, modifies the fields as the system evolves.
While this changes the detailed structures of the fields, it does not substantially modify field structures on scales beyond the coherence length of the field.

Finally, we touch on the properties of the two simulations
with magnetic fields internal to the bubbles (cases US-F and T$_L$S-F).
As introduced in \S 2.2, those calculations were initialized
with a tangled magnetic field contained within the bubble. The
maximum magnetic field just inside the bubble boundary
was approximately a factor of four greater than the average external field at a comparable ICM height.  As it
turns out, the internal field quickly dissipated as the bubbles
expanded due in part to weakening of the field from flux freezing,
but also in response to reconnection coming from the tangled nature of 
the original field geometry.
Consequently, the presence of internal magnetic fields was almost completely irrelevant to the evolution of our models.
Figure \ref{fig:tangleintrings} provides a comparison between the 
US and T$_L$S ({\it left}) and the US-F and T$_L$S-F ({\it right}) models.
There are subtle differences present in each pair, but the
internal fields clearly have not played significant roles.
We found a somewhat different consequence in our 2D simulations
\citep{jonesdeyoung05}. In those simulations circulations
established inside the bubbles stretched and amplified
the internal fields, which, however, were initially circumferential 
rather than tangled. This points to the importance of the field
geometry and to differences between 2D and 3D evolution.
In 3D  bubbles that were initially turbulent, so that turbulent
field amplification was ongoing, we might once again expect some
enhancement of the field so long as the turbulent motions
were strong enough to control the bubble field evolution on timescales
of order $\tau_{s,0}$.

\subsection{Bubble Upward Propagation}

\begin{figure*}[t]
\includegraphics[width=\textwidth]{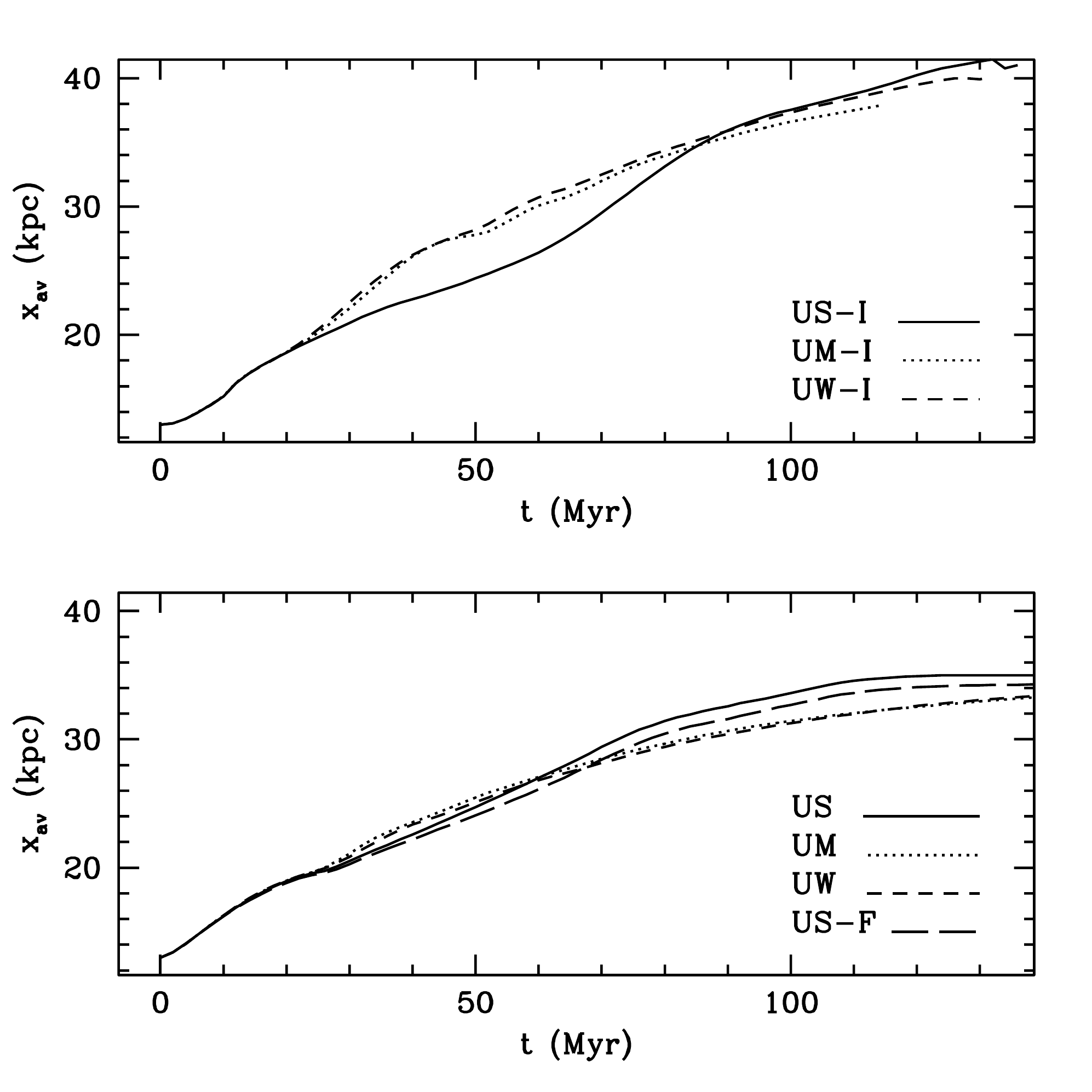}
\caption[Positions of the mean heights of bubbles in uniform ambient magnetic fields]{Positions of the mean heights of bubbles in uniform ambient magnetic fields. {\it Top:} Inflation models. {\it Bottom:} Models without inflation.}
%fig 12
\label{fig:upos}
\end{figure*}

Besides morphology the most applicable measure of bubble dynamics is 
a bubble's upward motion within the ICM. 
Figure \ref{fig:upos} illustrates this behavior for the
bubbles evolving in the quasi-uniform external fields. 
The quantity plotted is the mass-weighted mean elevation of the
bubbles, $x_{av} = \int C_f x \rho dV/\int C_f \rho dV$. 
As the discussion of the previous subsection suggests, the
upward motions of the bubbles placed in a tangled external
field are similar. The motions indicated in Figure \ref{fig:upos}
are close to those found for the analogous 2D bubbles
simulated by \citet{jonesdeyoung05}. They indicate an upward
velocity that is initially about $0.3c_{s,0}$, but which slowly
decreases over time, as the bubbles rise higher in the ICM. The
inflated bubbles decelerate somewhat less than the uninflated
bubbles.

The upward bubble motion is, of course, driven by buoyancy and limited by 
dynamical 
drag and Maxwell stresses. Along the way local dynamical features, 
such as R-T and K-H instabilities, as well as buoyancy-related
dynamical evolution of the bubble-ICM interface influence the
shape and integrity of the bubbles, as we have discussed. 
\citet{jonesdeyoung05} found
that the upward terminal velocity of their 2D 
bubbles were, in fact, well-described by the simple
balance between the buoyant force, $F_{buoy} \approx |g| \rho_0 V_b$,
and the drag force, $F_d \approx C_d A_b \rho_0 u^2_b$, where
$V_b$ and $A_b$ are the bubble volume and ``cross sectional' area, 
respectively, $u_b$ is the bulk speed of the bubble, 
$C_d \sim 1$ is the drag coefficient, and we utilized
the condition $\eta = \rho_b/\rho_0 \ll 1$.
Initially, when the bubble is quasi-spherical $V_b \approx (4/3)\pi r_b^3$
and $A_b \approx \pi r_b^2$, while after ring formation
$V_b \approx 2 \pi^2 R r_b^2$ and $A_b \approx 4\pi R r_b$,
where $R$ is the radius of the ring.
Balancing the buoyant and drag forces gives the familiar expression for the terminal upward
velocity,
\begin{equation}
u_t \sim \left (\frac{|g|V_b}{C_d A_b}\right )^{\frac{1}{2}}. 
\end{equation}
The factor $V_b/A_b$ is $(4/3)r_b$ for a sphere and $(\pi/2)r_b$
for a ring, so in either case the terminal velocity
can be written to sufficient accuracy simply as
\begin{equation}
u_t \approx c_{s,0}{\left (\frac{r_b}{h}\right )}^{\frac{1}{2}},
\label{uterm}
\end{equation}
where we set $\sqrt{V_b/(A_b C_d)} = 1$ and utilized the definition $h = c_{s,0}^2/|g|$.
The time to reach this equilibrium speed is very brief, about
\begin{equation}
\tau_t \sim \frac{\eta h}{c_{s,0}}{\left(\frac{r_b}{h}\right)}^{\frac{1}{2}} = \eta \tau_{s,0}{\left(\frac{r_b}{h}\right)}^{\frac{1}{2}} \ll  \tau_{s,0}.
\end{equation}

\citet{jonesdeyoung05} computed $V_b$ and $A_b$ numerically from
the simulation data, finding a close match between the actual upward
velocity and that predicted by equation \ref{uterm}. For our purposes
here a simpler analysis suffices.
Applying the initial bubble conditions to equation \ref{uterm} gives $u_t \approx 0.4 c_{s,0}$, which is reasonably close
to the measured velocities. The slow deceleration that
takes place at higher elevations can be understood from
the fact that the tube radius in the ring $r_b$, does not
increase substantially as the bubble rises. 
The radius of the torus,
$R$, does increase as the bubble rises, but that
has little impact on the terminal velocity,
which is proportional to $\sqrt{r_b/h}$.
Since $h$ increases in our simulations with elevation, the
terminal velocity 
is smaller at higher elevations.
The inflated bubbles are less decelerated than the uninflated
bubbles, because their tube radii are larger, as mentioned
before and as can be seen in the figures.  
In the first approximation the upward trajectories in Figure \ref{fig:upos}
are independent of the ambient magnetic field. There is, however, a notable dip in the
curves for the cases US, US-I and US-F beginning around $t \sim 30$ Myr. This
corresponds, of course, to the stage during which the ambient field bifurcates
those bubbles. 
Some of the bubble matter trails behind and actually begins to
fall, as a result of the reaction of the stretched field lines.
At the same time, however, bubble plumes continue to rise. The mass-weighted
bubble velocity decreases accordingly, but eventually recovers as the
plumes, which have characteristic sizes greater than the ring tube size
at the time of bifurcation, move
upward more rapidly than the unbifurcated rings.

\section{Conclusions}\label{sec:conc}

We conducted an ensemble of simulations designed to explore the 3D
nature of buoyant bubbles propagating in the magnetized ICM.

Our most important results are summarized here:

1. Consistent with previous 2D and 3D simulations we confirm
that magnetic fields of strength $\beta \sim 120$ can substantially 
affect the morphology of buoyantly rising bubbles. The influences
depend significantly on the geometry of the field.
In general, the results here do not show the extreme bubble break up
and subsequent mixing that are seen in the purely hydrodynamic
simulations of \citet{bruggenkaiser01} and \citet{bruggenetal02},
which is in accord with earlier MHD calculations.
In uniform horizontal fields, rising bubbles can become bifurcated as a
result of magnetic field trapping along the upper contact
discontinuity and subsequent stretching across the 
center of the bubble.
If the ambient field is tangled on scales smaller than the
size of the bubble, the dynamical influence of the field
is greatly reduced, since those fields are not
effectively amplified by the upward motions of the bubbles.
If the field tangling scale is significantly larger than
the bubble, the dynamical influence of the field can be
significant, although relative to a uniform field of
comparable strength the influence is retarded
and less coherent than that due to a uniform external field.

2. If the ambient fields have coherence scales larger than
the bubbles, the upward bubble motions can amplify the field
strengths locally and transport magnetic flux to higher elevations
in the ICM.
This field transport and amplification may lead to locally enhanced
mixing and lifting of the ambient ICM, and this process can in principle
contribute to the suppression of cooling flows.  However, the localized
nature of this process does not by itself provide an effective means of
raising the ICM temperature in the core on the global scales that are
needed.

3. The buoyant motions of bubbles can be approximated to zeroth order 
by a simple balance between buoyant and drag forces, but
the presence of magnetic fields and their consequent dynamical 
influence do have a clear effect upon the upward motion of the bubbles,
as is described in Section 3.3.

4. Hydrostatically formed bubbles, and in particular bubbles 
inflated slowly in the ICM, provide a useful tool for
understanding the  interactions of buoyant material with ambient
magnetic fields. In particular such models are applicable to the
evolution of the relic radio bubbles seen in the ICM of many rich
clusters.  In general, such spherical 
bubbles having scales comparable to the scale height of the local ICM
tend to produce toroidal bubble morphologies.  If, on the other hand,  
the objective is to understand the morphology of radio bubbles
directly inflated at the termination point of AGN jets, then
it is necessary to evolve that structure from
the dynamics of a jet rather than to artificially create a bubble
in hydrostatic equilibrium with the ICM. 

\acknowledgments
This work was supported at the University of Minnesota by NSF grant
AST06-07674 and by the University of Minnesota Supercomputing
Institute.  SMO further acknowledges the support of a Doctoral Dissertation Fellowship granted by the University of Minnesota.  The National Optical Astronomy Observatory is operated
by AURA, Inc., under a cooperative agreement with the
National Science Foundation.  We also thank the referee for their helpful comments.  Visualizations presented in this paper were constructed using the software packages FieldVis and The Hierarchical Volume Renderer (HVR), both developed at the University of Minnesota, and the commercial program Amira.

\clearpage

\appendix
\section{Acceleration of the Bottom Bubble-ICM Interface}
We outline two simple ways to estimate the early 1D acceleration
of the lower contact discontinuity, CD, between a buoyant bubble and its environment when the bubble is formed in an initially hydrostatic medium. Gravity points downwards in the $-x$ direction.
The motion of the CD is a response to a strong rarefaction that forms inside the bubble 
above the discontinuity. The rarefaction takes the ambient gas below the bubble
out of hydrostatic equilibrium, so that this gas accelerates upwards
towards the top of the bubble. 

The initial pressure distribution
everywhere is assumed to correspond to hydrostatic equilibrium in the ambient
medium; namely,
\begin{equation}
\frac{\partial P_0(x)}{\partial x} = \rho_0(x) g(x) < 0,
\label{pgrad}
\end{equation}
where $g(x)$ is the signed gravitational acceleration.
The initial CD is at $x = s$, with the bubble on top, so that $\rho(x>s) = \rho_b \ll \rho_0 = \rho(x<s)$.
As in the text we define $\eta = \rho_b/\rho_0\ll 1$.

The net acceleration of gas is
\begin{equation}
\ddot{x} = - \frac{1}{\rho} \frac{\partial P_0(x)}{\partial x} + g(x).
\label{accel}
\end{equation}
From equation \ref{pgrad} $\ddot{x} = 0$ initially for ambient gas, but not for the much
lighter bubble gas. The net acceleration of the bubble gas is, instead
\begin{equation}
\ddot{x} = -g(x)\left[\frac{\rho_0}{\rho_b} - 1\right] = - g(x)\frac{1 - \eta}{\eta}.
\label{baccel}
\end{equation}

The acceleration of the underlying ambient material occurs in response to
a rarefaction that
forms between the two substances, increasing the local pressure gradient
above the initial hydrostatic value.
The rarefaction propagates at the local sound speed, which is $c_{s,b}$ in the bubble and $c_{s,0}$ in the ambient medium.
Because of the initial pressure equilibrium, $c_{s,0} = \sqrt{\eta}~c_{s,b} \ll c_{s,b}$.
After a short time, $\Delta t$, the rarefaction moves into the bubble a
distance, $\Delta x_{r,b} = c_{s,b}\Delta t$, and oppositely into the ambient medium
a smaller distance, $\Delta x_{r,0} = - c_{s,0}\Delta t$. 
The pressure inside the rarefaction, $P_r$, is
approximately the initial pressure just above the rarefaction, so 
$P_r \approx P(x = s + \Delta x_{r,b})$. Using equation
\ref{pgrad} this gives $P_r \approx P(s) + \rho_0~g\Delta x_{r,b}$. The pressure difference
across the rarefaction propagating into the ambient medium then becomes approximately
$\Delta P \approx -  \rho_0~g(\Delta x_{r,b} - \Delta x_{r,0})$
Using equation \ref{accel}, the net acceleration of ambient gas in the rarefaction is then approximately
\begin{equation}
\begin{aligned}
\ddot{x} &= \frac{1}{\rho_0} \left[\frac{- \rho_0 g(x) (\Delta x_{r,b}-\Delta x_{r,0})}{|\Delta x_{r,0}|}\right] + g(x)\\
~~~ &= g(x)\left[\frac{\Delta x_{r,b}}{\Delta x_{r,0}}\right].
\end{aligned}
\end{equation}
The ratio $\Delta x_{r,b}/\Delta x_{r,0}$ is simply the negatively signed ratio of the sound speeds, 
$-c_{s,b}/c_{s,0}$, so associating this motion with the CD gives
\begin{equation}
\ddot{x}_{CD} = -g(x)\left[\frac{c_{s,b}}{c_{s,0}}\right] = - g(x) \frac{1}{\eta^{\frac{1}{2}}}.\\
\label{cdone}
\end{equation}

Alternatively, one can estimate the motion of the CD from the solution of an approximate Riemann
problem representing the initial conditions.  Ignoring gravity for now
and setting $\gamma = 5/3$, the Riemann
invariant, $J_- = u - 3c_{s,b} = {\rm constant}$, applies inside the rarefaction propagating into the
bubble from its lower boundary, while inside the associated rarefaction 
moving into the ambient medium $J_+ = u + 3c_{s,0} = {\rm constant}$ applies. 
The local velocity is given by $u$. Just above the rarefaction penetrating
the bubble the velocity is, according to equation \ref{baccel},
$u_b \approx -g  \Delta t(1 - \eta)/\eta$, whereas just below
the rarefaction penetrating into the ambient medium $u_0 = 0$.
The Riemann invariants allow us to find the matching conditions
at the CD, and specifically its velocity,
\begin{equation}
\begin{aligned}
u_{CD} &= u_0 + c_{s,B}\left[\frac{u_0 - u_b}{c_{s,0}+c_{s,b}}\right] 
&= - \frac{g \Delta t}{\eta^{\frac{1}{2}}}
\frac{1 - \eta}{1 + \eta^{\frac{1}{2}}}.\\
\end{aligned}
\end{equation}
Since $\eta \ll 1$, we have once again (letting $\ddot{x}_{CD} \approx u_{CD}/\Delta t$)
\begin{equation}
\ddot{x}_{CD} \approx - \frac{g}{\eta^{\frac{1}{2}}}.
\label{cdtwo}
\end{equation}
We have verified that this behavior is consistent with 1D simulations of
a contact discontinuity set up with these initial conditions.

\bibliography{refs}

\end{document}